\documentclass[twocolumn,showpacs,aps,prd,amssymb,nofootinbib,nobibnotes,floatfix,longbibliography]{aastex631}

\usepackage{newtxtext,newtxmath}
\usepackage{ae,aecompl}
\usepackage[mathscr]{euscript}

\usepackage[T1]{fontenc}
\usepackage{hyperref}
\usepackage{amsfonts}
\usepackage{amsmath}
\usepackage{mathrsfs}
\usepackage{epsfig}
\usepackage{graphicx}
\usepackage{url}
\usepackage{hyperref}
\usepackage{float}
\usepackage{color}
\usepackage[utf8]{inputenc} 
\usepackage{epsf}
\usepackage{comment}
\usepackage{slashed}
\usepackage{footmisc}
\usepackage{lipsum}
\definecolor{linkcolor}{rgb}{0.0,0.3,0.5}
\usepackage[caption=false]{subfig}

\usepackage{booktabs}

\usepackage{color}
\definecolor{cerulean}{rgb}{0.0, 0.48, 0.65}

\definecolor{navy}{rgb}{0.2, 0.0, 1.0}

\definecolor{jungle}{rgb}{0.0, 0.5, 0.0}

\usepackage{color}

\definecolor{orange}{rgb}{1,0.5,0}
\definecolor{orangeB}{rgb}{1,0.7,0}

\usepackage{ulem}
\usepackage{units}
\usepackage{soul}
\usepackage{orcidlink}

\usepackage[mathlines]{lineno}

\begin{document}

\preprint{APS/123-QED}

\title{The effect of pressure confinement on the maximum mass of rapidly accreting supermassive stars}

\author{Lorenz Zwick}
\affiliation{Niels Bohr International Academy, The Niels Bohr Institute, Blegdamsvej 17, DK-2100, Copenhagen, Denmark}
\affiliation{Center of Gravity, Niels Bohr Institute, Blegdamsvej 17, 2100 Copenhagen, Denmark.}
\email{lorenz.zwick@nbi.ku.dk}

\author{Christopher Tiede}
\affiliation{Niels Bohr International Academy, The Niels Bohr Institute, Blegdamsvej 17, DK-2100, Copenhagen, Denmark}

\author[0000-0003-3633-5403]{Zolt\'an Haiman}
\affiliation{Institute of Science and Technology Austria (ISTA), Am Campus 1, Klosterneuburg 3400 Austria}

\shorttitle{GR instability in pressure confined SMS}
\shortauthors{Zwick et al.}

\date{\today}

\begin{abstract}
\noindent
We derive an extension of Chandrasekhar's pulsational stability criterion for relativistic stars with a pressure boundary condition at the surface. We then apply the new criterion to hylotropes, a model describing rapidly accreting supermassive stars with a core-envelope structure. We find that confinement typically stabilises stars, delaying the onset of the general relativistic instability when the external pressure at the surface is roughly $\sim10^{-3}$ of the stars' hydrostatic pressure scale. Beyond this value, the critical mass at marginal stability increases rapidly, exceeding the value of $\sim 10^5$ M$_{\odot}$ for helium burning, isolated $n=3$ polytropes by orders of magnitude. We discuss how the required pressure boundary conditions may be produced by the combined ram and magnetic pressure supplied by the accretion flow as it assembles the supermassive star, and provide a fitting formula for the final mass before collapse. As a rough estimate assuming equipartition between magnetic and thermal pressures in a standard $\alpha$-disk, we find that the critical mass doubles at accretion rates of order $\dot M\sim100\,{\rm M_\odot}\,{\rm yr}^{-1}$, again increasing rapidly after this value. Our results show under what conditions, potentially realized in a subset of high redshift halos, non-rotating SMS may collapse into black holes beyond the typical scale of heavy seeds.
\end{abstract}

\section{Introduction}
\label{S:Introduction}

The formation and growth of supermassive black holes (BH) at high redshift remains one of the outstanding problems in astrophysics 
\citep{Volonteri:2012kq,2013Haiman,Woods2019,InayoshiVisbalHaiman2020,2024OJAp....7E..72R}. Broadly, proposed formation pathways have been classified into two categories: light seeds that subsequently undergo phases of rapid growth \citep{Madaurees2001,Abel2002,Schneider2002,Hirano2014,2026NatAs..10..583M}, and intrinsically heavy seeds produced through some form of direct collapse \citep{OhHaiman2002,BrommLoeb:2003,LodatoNatarajan:2006,BegelmanVolonteriRees:2006,2008begelman,2008MNRAS.387.1649B,2019Mayer,2020regan,2024NatAs...8..126B}. The former requires seed formation to occur extremely early (z $\gtrsim$ 20) as well as extended periods of super-Eddington accretion \citep{TanakaHaiman2009,VolonteriRees2006,Zhu2020,Sassano2021,2024MNRAS.527.1033B}, while the latter requires the assembly and coherent collapse of a sufficiently massive, low-angular-momentum gas reservoir \citep{Colgate2003,2004koushippas,2007lodato,2013prieto,2016agarwal,2022latif,2024A&A...692A.213P}. Both classes of models have advantages and limitations, and both can be made consistent with the observed quasar luminosity function under suitable assumptions \citep{LippaiFreiHaiman2009,2018MNRAS.481.3278R,2023Spinoso,2023MNRAS.519.4753T}. Because of the fact that exponential BH growth erases the initial condition, it is challenging to find a clear observational preference for either model \citep[e.g.][]{2024MNRAS.531.4584S}. In any case, a basic underlying requirement of rapid black hole growth is that large amounts of gas, assembled from galactic scales of tens to thousands of parsecs, must ultimately reach radii of order the innermost stable circular orbit. Crucially, the latter is only $\sim 3\times10^{-8}\,{\rm pc}$ for a non-rotating $10^5\,{\rm {\rm M_\odot}}$ black hole. The major obstacle is therefore the removal of angular momentum required to maintain a coherent inflow across more than ten orders of magnitude in spatial scale.

This same requirement to grow BHs in the centre of high redshift halos favours the formation of a different type of object, i.e. of supermassive stars (SMS). While their existence is not confirmed, SMSs have long been considered a possibility \citep{BondArnettCarr1984,FullerWoosleyWeaver1986}, 
and have been proposed as promising progenitors of massive black-hole seeds \citep{begelman2010,2012hosokawa,hosokawa2013,haemmerle2018a,2023herrington,2025lionel}. The main argument in their favour is precisely their scale. Highly accreting SMSs are expected to evolve approximately along the Hayashi limit as cool, bloated proto-stars with effective temperatures of order $5000\,{\rm K}$ \citep{1961hayashi,hosokawa2013}. According to this condition, a $10^5\,{\rm M_\odot}$ SMS will have a radius of order $200\,{\rm AU}\simeq10^{-3}\,{\rm pc}$, more than four orders of magnitude larger than the ISCO of a black hole of the same mass. For this reason, it should be significantly easier to sustain an accretion flow onto an SMS rather than directly onto a black hole. Additionally, because accretion terminates in a much shallower region of the gravitational potential, the radiative efficiency from accretion is much lower than from a BH with the same mass.
Feedback is less efficient at shutting off the inflow and large accretion rates can be more easily sustained \citep{1921eddington,hosokawa2013,chon2018,2024kiyuna}. So, while SMSs remain theoretical objects, these arguments strongly suggest that they should form under the same conditions required to produce and grow black-hole seeds of essentially any kind. Intriguingly, the discovery of ``Little Red Dots'' by \textit{JWST} \citep{Matthee+2024-LRD,Kocevski:LRDs:2023,Labbe:LRDbalmerbreak:2024,2024setton,akins2024,perez2024,2024killi,2024yue,2025furtak}, has renewed interest in rapidly accreting SMSs, as well as related BH-star objects, as their cool and extended atmospheres offer an explanation for their red continua, broad Balmer emission, and weak X-ray luminosities \citep{NaiduMatthee:2025,2026ApJ...998..124N,2026ApJ...996...48B,2026ApJ..1002....7Z,2026arXiv260321714R}.

For sufficiently high accretion rates, the final fate of a SMS is determined by the onset of the General-Relativistic (GR) instability, first formally derived by Chandrasekhar in 1964 \cite{1964chandra}. This instability arises because relativistic corrections modify the response of the star to dynamical perturbations about hydrostatic equilibrium. In sufficiently massive, radiation-pressure-dominated stars, these relativistic corrections can globally destabilise the star against radial perturbations, an effect that is entirely absent in the corresponding Newtonian treatment. Once the star becomes unstable, it is expected to undergo coherent collapse and form a massive black-hole seed with relatively little mass loss \citep{2002ApJ...572L..39S,2009saijo,2013reisswig,2016PhRvD..94b1501S}. For isolated SMSs, adopting Chandrasekhar's criterion results in the instability typically setting in at masses of $\sim10^5\,\rm{M}_\odot$, although the precise threshold depends on the accretion rate, the internal structure and the composition of the star \citep{hosokawa2013,2020lionel,2021woods,2025arXiv251108516N}. Importantly, additional sources of support, in particular rotation, can delay the onset of the instability and allow the star to reach significantly larger masses before collapse \citep{baumgarte1999b,shibata2016a,2021lionel,haemmerle2018b}, potentially providing exceptionally heavy BH seeds required to explain outliers in the quasar mass function such as J0100+2802, with an estimated black-hole mass of $\sim10^{10}\,{\rm M_\odot}$ at $z=6.30$, and J0313$-$1806, hosting a $\sim1.6\times10^9\,{\rm M_\odot}$ black hole already at $z=7.64$ \citep{2015Natur.518..512W,Wang2021, 2023ApJ...950...68E}.

In almost all applications of the GR instability, the star is effectively treated as an isolated system. In terms of the derivation of the instability criterion, this corresponds to imposing a vanishing pressure boundary condition at the stellar surface, as in the original work by Chandrasekhar. However, this assumption is difficult to reconcile with realistic formation of SMSs at high redshifts, which requires the existence of a large-scale accretion flows spanning the range of $\sim0.1-10^3\,{\rm M_\odot}\,{\rm yr}^{-1}$, from galactic scales toward the centre of the halo \citep{hosokawa2013,haemmerle2018a,2020regan,2023herrington,2024kiyuna,2024A&A...689A.169S}. The inflow required to build and sustain the star will exert ram pressure and may also advect or generate dynamically important magnetic fields \citep{2013Latif,2022latif,haemmerle2019a,2023ApJ...952L...9L}. Rather than the star being isolated, we should therefore expect the SMS to be confined by a finite external pressure. 

Motivated by this, here we revisit the original derivation of the GR instability and extend it to allow for a pressure-confined boundary condition. We derive a general modification of the stability criterion, and consider ram pressure and magnetic confinement as realistic examples. We then apply the revised criterion to hylotropic stellar models, which
provide an accurate analytical description of rapidly accreting SMSs over accretion rates from $\sim0.1$ to $\gtrsim10^3\,{\rm M_\odot}\,{\rm yr}^{-1}$, encompassing the range most relevant for black-hole seed formation in high-redshift halos \citep{hosokawa2013,haemmerle2018a,haemmerle2019c,2021lionel,2025lionel}. We find that pressure confinement generally delays the onset of the GR instability. Under certain conditions, such as a very high mass accretion rate or magnetic pressure, this allows the SMS to grow to larger masses before collapse, resulting in higher maximum luminosities and also increasing the maximum mass of the resulting black-hole seed.
\newline
\newline
The paper is organised as follows. In Section~\ref{S:GR_Instability}, we review the GR instability of radiation-dominated stars and derive the modification to Chandrasekhar's stability criterion introduced by a finite pressure boundary condition, both heuristically and rigorously. In Section~\ref{S:hylotropes}, we apply the revised criterion to hylotropic models of rapidly accreting SMSs, determine the corresponding stability boundaries, discuss realistic sources of pressure confinement in high-redshift accretion flows, and derive the corresponding critical masses. In Section~\ref{S:discussion}, we discuss the astrophysical implications of our results, as well as present some broader concluding remarks.

\section{General-relativistic instability in pressure confined stars}
\label{S:GR_Instability}

\subsection{Background and heuristic derivation}
\label{S:GR_Instability:heuristic}
The evolution of a rapidly accreting SMS proceeds roughly through the following stages. Initially, the growth of the star is dominated by accretion. The stellar structure is not yet thermally relaxed, and the SMS evolves as a cool, bloated protostar with a radius set approximately by the Hayashi limit \citep{2012hosokawa,hosokawa2013,haemmerle2018a,2023herrington}. Crucially, massive stars are close to being entirely supported by radiation pressure \citep{Kippenhahn:1994tm}. For a radiation-pressure-dominated gas, the first adiabatic index may be written as:
\begin{equation}
\gamma_*
\simeq
\frac{4}{3}
+
\frac{\beta_*}{6},
\label{eq:gas_pressure_gamma_correction}
\end{equation}
where:
\begin{equation}
\beta_*
\equiv
\frac{P_*^{\rm gas}}{P_*}
\propto
\rho_* T_*^{-3}
\label{eq:beta_definition}
\end{equation}
is the ratio of gas pressure to total pressure. Here the subscript $*$ denotes typical or scale values for the entire star, regardless of its internal structure. A pure radiation gas has $\beta_*=0$ and therefore $\gamma_*=4/3$ exactly. It is a classic result that a star in Newtonian gravity is dynamically stable against radial perturbations when its first adiabatic index is larger than a critical value \citep{1898ApJ.....8..293R}:
\begin{equation}
\gamma_*
>
\gamma_{\rm crit}^{\rm Newton}
\equiv
\frac{4}{3}.
\label{eq:newtonian_gamma_critical}
\end{equation}
Together with Eq.~\eqref{eq:gas_pressure_gamma_correction}, this shows that an adiabatic mixture of gas and radiation is always dynamically stable in Newtonian gravity, while a pure radiation gas is precisely marginal. Chandrasekhar famously showed that GR shifts this threshold. The full calculation involves several steps, which we detail in Section~\ref{S:gr_stability}. However, for a weakly relativistic star with a polytropic equation of state, the entire stability criterion reduces to a shift in the critical adiabatic index which may be written as:
\begin{equation}
\gamma_{\rm crit}^{\rm GR}
\simeq
\frac{4}{3}
+
\kappa_{\rm GR}
\frac{GM_*}{R_*c^2},
\label{eq:gr_gamma_critical}
\end{equation}
where $M_*$ and $R_*$ are characteristic stellar mass and radius, and $\kappa_{\rm GR}$ is a dimensionless coefficient determined by the stellar structure. GR therefore raises the critical adiabatic index and destabilises the star. As an SMS grows in mass, an increasing fraction of its interior can thermally relax, become hotter, and shift the effective adiabatic index closer to $4/3$. At the same time, the stellar compactness increases, making the relativistic correction larger. The onset of the GR instability is then precisely determined when the stabilising gas-pressure correction and the destabilising relativistic correction become comparable:
\begin{equation}
\frac{\beta_*}{6}
\simeq
\kappa_{\rm GR}
\frac{GM_*}{R_*c^2},
\label{eq:standard_gr_marginal_condition}
\end{equation}
after which the SMS collapses to a BH in roughly a dynamical timescale retaining the vast majority of its mass \citep[see][for a numerical GR validation, also in the presence of rotation and magnetic fields within the stellar interior]{2018sun}.

A crucial assumption in Chandrasekhar's original formulation, and in the vast majority of subsequent work, is that the stability analysis is performed with a vanishing pressure boundary condition at the stellar surface. Here, we present a simple scaling argument to illustrate why external pressure confinement modifies the stability condition. As before, we consider an SMS characterised by a scale radius $R_*$, mass $M_*$, and internal pressure scale $P_*$ in hydrostatic equilibrium. Hydrostatic equilibrium represents a virial balance between the internal and gravitational energy:
\begin{equation}
\underbrace{P_*V_*}_{\text{intern.}}
\sim
\underbrace{\frac{GM_*^2}{R_*}}_{\text{grav.}},
\label{eq:heuristic_virial_vacuum}
\end{equation}
where $\sim$ denotes commensurability, and $V_*\sim R_*^3$ is the characteristic stellar volume. We now ask whether this equilibrium is stable under a homologous contraction of the SMS at fixed mass. For an adiabatic equation of state:
\begin{equation}
P_*
\propto
\rho_*^{\gamma_*}
\propto
V_*^{-\gamma_*},
\end{equation}
and therefore the two sides of Eq. \ref{eq:heuristic_virial_vacuum} scale as:
\begin{equation}
P_*V_*
\propto
V_*^{1-\gamma_*}
\propto
R_*^{3-3\gamma_*},
\qquad
\frac{GM_*^2}{R_*}
\propto
R_*^{-1}.
\label{eq:heuristic_vacuum_scalings}
\end{equation}
We see that the internal-energy, responsible to sustain the star against its own self-gravity, reacts more ``stiffly'' to contraction than the gravitational energy when:
\begin{equation}
3-3\gamma_*<-1.
\end{equation}
Equivalently, when:
\begin{equation}
\gamma_*>\frac{4}{3},
\end{equation}
which is, in fact, precisely the familiar Newtonian stability threshold for the stability of stars. We now include an external pressure $P_{\rm ext}$ acting at the stellar surface. We characterize the response of the external pressure to a displacement of the stellar surface by:
\begin{equation}
s
\equiv
-\frac{d\ln P_{\rm ext}(R_*)}{d\ln R_*}.
\label{eq:external_pressure_response}
\end{equation}
such that the pressure applied at the stellar boundary is:
\begin{equation}
P_{\rm ext}(R_*)
\propto
R_*^{-s}.
\end{equation}
We can repeat the virial-balance argument while accounting for an additional source of pressure confinement at the surface of the star. In analogy with the distinction between the Jeans and Bonnor--Ebert problems and as studied in \citet{1970MNRAS.151...81H}, the presence of a finite external pressure modifies the virial balance of the star. The virial equilibrium condition becomes \citep{1902RSPTA.199....1J,1955ZA.....37..217E,1956MNRAS.116..351B}:
\begin{equation}
\left(P_*-P_{\rm ext}\right)V_*
\sim
\frac{GM_*^2}{R_*},
\label{eq:schematic_virial}
\end{equation}

To determine how this modifies the stability criterion, we again compare the radial scalings of the different terms: The external-pressure contribution scales as:
\begin{equation}
P_{\rm ext}V_*
\propto
R_*^{3-s},
\end{equation}
whereas the gravitational term scales as $R_*^{-1}$. Therefore, the contribution associated with external confinement is stabilising when $4-s>0$ and destabilising when $4-s<0$. In the limit $P_{\rm ext}\ll P_*$, we therefore expect the critical adiabatic index to be modified heuristically as:
\begin{equation}
\gamma_{\rm crit}^{\rm ext}
\simeq
\frac{4}{3}
+
\kappa_{\rm GR}
\frac{GM_*}{R_*c^2}
-
\kappa_{\rm ext}(4-s)
\frac{P_{\rm ext}}{P_*},
\label{eq:heuristic_gamma_critical}
\end{equation}
where we added the relativistic and pressure corrections independently. Here, the dimensionless coefficient $\kappa_{\rm ext}$ absorbs the numerical factors neglected in the scaling argument.

Eq.~\eqref{eq:heuristic_gamma_critical} already captures the main physical result of this work. For $s<4$, external confinement lowers the critical adiabatic index and is therefore stabilising. The corresponding marginal-stability condition becomes:
\begin{equation}
\frac{\beta_*}{6}
\simeq
\kappa_{\rm GR}
\frac{GM_*}{R_*c^2}
-
\kappa_{\rm ext}(4-s)
\frac{P_{\rm ext}}{P_*}.
\label{eq:heuristic_marginal_condition}
\end{equation}
Crucially, this shows that the stability of a rapidly accreting SMS is controlled by two small and competing corrections: GR destabilises the star, while external pressure provides an additional stabilising contribution, provided that $s<4$. This additional external confinement can therefore delay the onset of collapse by allowing the star to reach a larger relativistic correction before becoming unstable. The confinement term becomes dynamically relevant once $P_{\rm ext}/P_*$ is comparable to the relativistic correction, namely to the compactness of the SMS. Both the hylotropic model adopted in Section~\ref{S:hylotropes} and stellar-evolution calculations consistently find that the GR instability is reached for relativistic corrections of order $10^{-3}$. We therefore expect the two effects to become competitive when:
\begin{equation}
P_{\rm ext}
\sim
10^{-3}\,P_*.
\label{eq:heuristic_pressure_threshold}
\end{equation}
This represents only a small fraction of the characteristic stellar pressure, suggesting that the required confinement may be realised in plausible astrophysical environments. It is important to note the correspondence of this result to the effect of rotational support, which can delay the GR instability even when the star's rotational energy is only a small fraction of the gravitational energy \citep{shibata2016a,2021lionel}, in analogy with the heuristic argument presented here. Additionally, it is also similar to stability calculations which considered additional media, such as a dark matter component \citep{2024PhRvD.110h3035K} or magnetic fields \citep{2022MNRAS.516.1481L}. In all of these cases, the large response to a comparatively small correction ultimately reflects the peculiar marginality of a radiation pressure dominated star in Newtonian gravity: a pure photon gas has $\gamma_*=4/3$ and therefore lies exactly at the threshold of dynamical stability.\footnote{Following a private communication with I.~Linial, we note that the stability of stars appears to be a special property of $d=3$ spatial dimensions, due to the required balance between degrees of freedom for monoatomic gases and Gauss' law for gravity. In fact, it is possible to conclude that the observation of massive stars such as R136a1 \citep{2010MNRAS.408..731C} roughly constrains the dimensionality of space to be $d\lesssim3.2$. Not the finest of tuning arguments, but nonetheless something to be thankful for.}

\subsection{Extension of the Chandrasekhar stability criterion}
\label{S:gr_stability}
To extend the GR instability criterion to a pressure-confined star, we follow closely the original derivation of Chandrasekhar and the reformulation introduced in \cite{2020lionel}. The latter provides a particularly convenient rearrangement of Chandrasekhar's variational integrals, in which the Newtonian stability threshold at $\gamma=4/3$ is cleanly separated from relativistic corrections \citep[though see also][for a different approach]{2022MNRAS.517.1584N}. The full calculation is algebraically involved, and we present a self-contained derivation in Appendix~\ref{S:derivations}. Here, we summarise only the main steps and the origin of the additional pressure confinement term. We note that both Chandrasekhar's derivation and our extension rely on three fundamental assumptions: (i) spherical symmetry, (ii) conservation of baryon number, (iii) linear perturbations about relativistic hydrostatic equilibrium, and (iv) homologous trial displacements. We additionally introduce the assumptions that: (v) the medium responsible for the pressure confinement does not significantly perturb the background metric, and (vi) the response of the external pressure to stellar pulsations can be entirely described by a profile $P_{\rm ext}(R_*)$. We will return to discussing these assumptions in Section~\ref{S:discussion}. We note also that different approaches to extend the GR instability criterion exist, though to our knowledge the specific problem of pressure boundary conditions has not appeared in the literature \citep[as opposed to additional pressure sources within the star, as in e.g.][]{2018MNRAS.477.3694B}. 

Our starting point is the spherically symmetric, static metric:
\begin{equation}
    ds^2
    =
    -e^{2a(r)}(c\,dt)^2
    +
    e^{2b(r)}dr^2
    +
    r^2d\Omega^2,
\end{equation}
and Chandrasekhar's radial pulsation equation for a relativistic star. The equation is cast as a variational problem for the radial Lagrangian displacement $\xi(r)$, which requires an integration by parts. For an isolated SMS, the equilibrium pressure vanishes at the stellar surface:
\begin{equation}
P(R_*)=0,
\end{equation}
and the Lagrangian pressure perturbation also vanishes,
\begin{equation}
\Delta P(R_*)=0.
\end{equation}
These conditions remove all endpoint contributions from the integration by parts. Instead, for a pressure-confined star we must retain a boundary term, which we denote as $B_{\rm C}$. Explicitly, we find:
\begin{equation}
\mathcal B_{\rm C}
=
R_*^2 e^{a_s+b_s}\xi_s\Delta P_s,
\label{eq:main_general_surface_term}
\end{equation}
where $\Delta P_s$ is the Lagrangian pressure perturbation, and the subscript $s$ denotes quantities evaluated at the stellar surface. This boundary term represents the work performed as the stellar surface is displaced. Again, we characterize the instantaneous response of the external pressure by Eq. \ref{eq:external_pressure_response}. Then, a displacement of the stellar surface $\Delta R_*=\xi_s$ gives:
\begin{equation}
\Delta P_{\rm ext}
=
-sP_{\rm ext}\frac{\xi_s}{R_*},
\end{equation}
and therefore:
\begin{equation}
\mathcal B_{\rm C}
=
-sP_{\rm ext}R_*e^{a_s+b_s}\xi_s^2.
\label{eq:main_chandrasekhar_boundary}
\end{equation}
Following Chandrasekhar, we specialise the variational problem to relativistic homologous displacements:
\begin{equation}
\xi(r)
=
r e^{a(r)}.
\label{eq:main_homologous_displacement}
\end{equation}
For this choice:
\begin{equation}
\mathcal B_{\rm C}
=
-se^{3a_s+b_s}P_{\rm ext}R_*^3.
\label{eq:main_chandrasekhar_boundary_homologous}
\end{equation}
We then follow the rearrangement introduced in \cite{2020lionel} in order to isolate the relativistic correction to the Newtonian marginal value $\gamma=4/3$. This requires an additional integration by parts, whose endpoint contribution vanishes for an isolated star. In the pressure-confined case, it instead generates the additional boundary term:
\begin{equation}
\mathcal B_{\rm H}
=
4e^{3a_s+b_s}P_{\rm ext}R_*^3.
\label{eq:main_haemmerle_boundary}
\end{equation}
Collecting all contributions, the marginal-stability condition for a homologous displacement can be written as:
\begin{equation}
0
=
I_1+I_2+\mathcal S_3+\mathcal S_4
+\mathcal B_{\rm C}
+\mathcal B_{\rm H},
\label{eq:main_S12_rearranged}
\end{equation}
where:
\begin{align}
I_1
&=
9\int_0^{R_*}
e^{3a+b}
\left(
\gamma-\frac{4}{3}
\right)
Pr^2\,dr,
\\
I_2
&=
-12\int_0^{R_*}
e^{3a+b}
\left(
\frac{da}{dr}
+
\frac{1}{3}\frac{db}{dr}
\right)
Pr^3\,dr,
\\
\mathcal S_3
&=
\frac{8\pi G}{c^4}
\int_0^{R_*}
e^{3b+a}
P(P+\epsilon)r^2\xi^2\,dr,
\\
\mathcal S_4
&=
-\int_0^{R_*}
e^{a+b}
\frac{1}{P+\epsilon}
\left(
\frac{dP}{dr}
\right)^2
r^2\xi^2\,dr.
\end{align}
Here $P(r)$ and $\epsilon(r)$ are the equilibrium pressure and total energy-density profiles, respectively. Crucially, the two boundary terms combine into:
\begin{equation}
\mathcal B_{\rm C}
+
\mathcal B_{\rm H}
=
(4-s)e^{3a_s+b_s}P_{\rm ext}R_*^3,
\label{eq:main_combined_boundary}
\end{equation}
so that the marginal-stability condition becomes:
\begin{equation}
0
=
I_1+I_2+\mathcal S_3+\mathcal S_4
+
(4-s)e^{3a_s+b_s}P_{\rm ext}R_*^3.
\label{eq:main_dimensional_stability_condition}
\end{equation}
The advantage of this form is that $I_1$ contains the departure from the Newtonian marginal value $\gamma=4/3$, while $I_2$, $\mathcal S_3$, and $\mathcal S_4$ contain the relativistic corrections, as will become apparent after expanding the equilibrium quantities in powers of $1/c^2$. Importantly, the same factor $4-s$ obtained in the heuristic argument appears in the full variational calculation.

We now expand the stability condition simultaneously near the Eddington limit, where the star is almost entirely supported by radiation pressure, and to first post-Newtonian order:
\begin{equation}
\gamma-\frac{4}{3}
=
\frac{\beta}{6}
+
\mathcal O(\beta^2),
\qquad
\beta
\equiv
\frac{P_{\rm gas}}{P}.
\end{equation}
The variational integrals then simplify to:
\begin{align}
0
={}&
\frac{3}{2}
\int_0^{R_*}
\beta P r^2\,dr
-
\frac{8G}{c^2}
\int_0^{R_*}
PM_r r\,dr
\nonumber\\
&-
\frac{8\pi G}{c^2}
\int_0^{R_*}
P\rho r^4\,dr
-
\frac{G^2}{c^2}
\int_0^{R_*}
M_r^2\rho\,dr
\nonumber\\
&+
\underbrace{(4-s)P_sR_*^3
\left(
1-\frac{2GM_*}{R_*c^2}
\right)}_{\text{pressure confinement term}}\,
+\,
\mathcal O\!\left(
\beta^2,
\beta \mathcal C,
\mathcal C^2
\right),
\label{eq:main_dimensional_PN_stability_condition}
\end{align}
where:
\begin{equation}
\mathcal C
\equiv
\frac{GM_r}{rc^2}
\label{eq:stellar_compactness}
\end{equation}
is the compactness of the enclosed mass $M_r$. The first term in Eq. \ref{eq:main_dimensional_PN_stability_condition} describes stabilisation by gas pressure, the following three terms contain the leading relativistic destabilisation, and the final term is the contribution from pressure confinement. Gas pressure raises the effective adiabatic response above $\gamma=4/3$, GR destabilises the star at post-Newtonian order, and external confinement contributes an additional term proportional to $4-s$. Quite pleasantly, we find that Eq.~\eqref{eq:main_dimensional_PN_stability_condition} matches the heuristic stability condition derived in Section~\ref{S:GR_Instability:heuristic}, now integrated over the full hydrostatic structure of the star. However, to quote \citet{1964chandra} on such heuristic derivations:
\newline
\newline
``\textit{While such arguments are physically plausible (and sometimes give the correct results), it is known that they are not always conclusive. In any event, it is clear that the question of the dynamical stability of a gaseous mass can be answered without any ambiguity by an analysis of its normal modes of radial oscillation.}''

\section{Application to hylotropic SMS}
\label{S:hylotropes}
\begin{figure}
    \centering
    \includegraphics[width=0.8\columnwidth]{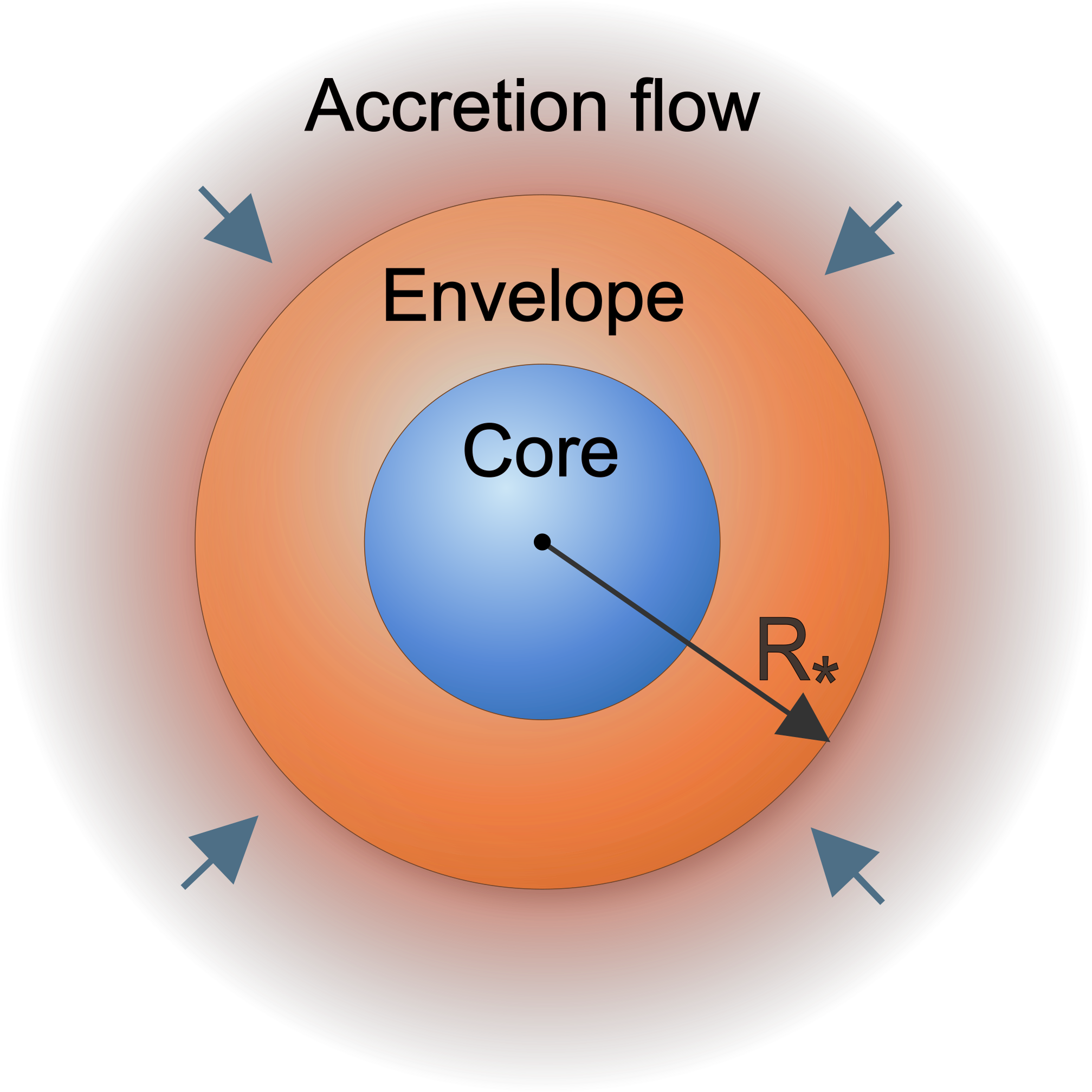}
    \caption{Simple sketch of the structure of a rapidly accreting supermassive star. A thermally relaxed, isentropic core is surrounded by an extended envelope. The ratio of the core to envelope mass depends strongly on the accretion rate, with larger envelopes requiring rapid accretion. In addition to assembling the star itself, the accretion flow can provide external pressure confinement through ram pressure or the advection of magnetic fields. This additional boundary condition modifies the response of the star to radial perturbations.}
    \label{fig:sketch}
\end{figure}

\subsection{Pressure confined hylotropic models }
\label{S:hylotropic_model}

The stability criterion derived in Section~\ref{S:gr_stability} applies generally to weakly relativistic, radiation-dominated stars subject to an external pressure. As discussed in Section~\ref{S:GR_Instability:heuristic}, a rapidly accreting SMS initially grows as a bloated protostar with an extended envelope. Once the thermal timescale becomes sufficiently short in the inner regions, the stellar core begins to relax, contract, and eventually ignite nuclear fusion. Following \cite{begelman2010} and \cite{2020lionel}, we describe the resulting structure as a hylotrope, a physically motivated model employing an equation state that generalises polytropes to allow an explicit dependence on the enclosed mass.
These models accurately reproduce the results of stellar evolution codes. Here, we briefly review the basic properties of hylotropic stellar models before applying the modified stability criterion derived above.

In the hylotropic model for SMS, the star is divided into an isentropic core and an envelope whose entropy increases outwards (see Fig. \ref{fig:sketch} for an illustration). The core obeys an approximately adiabatic equation of state with $\gamma_{\rm core}\simeq4/3$, whereas the envelope accrues mass homologously due the rapid accretion of new layers. The pressure and gas-pressure fraction are written as:
\begin{align}
P
&=
\begin{cases}
K\rho^{4/3},
    & M_r\leq M_{\rm core}, \\[3pt]
K\left(\dfrac{M_r}{M_{\rm core}}\right)^{2/3}\rho^{4/3},
    & M_r>M_{\rm core},
\end{cases}
\label{eq:hylotrope_eos}
\\
\beta
\equiv
\frac{P_{\rm gas}}{P}
&=
\begin{cases}
\beta_c,
    & M_r\leq M_{\rm core}, \\[3pt]
\beta_c\left(\dfrac{M_r}{M_{\rm core}}\right)^{-1/2},
    & M_r>M_{\rm core},
\end{cases}
\label{eq:hylotrope_beta}
\end{align}
where $M_r$ is the mass enclosed within radius $r$, $M_{\rm core}$ is the mass of the thermally relaxed core, and $\beta_c$ is the gas-pressure fraction within the core.
If the core contains the entire stellar mass, $M_{\rm core}=M_*$, the model reduces to an $n=3$ polytrope. As the fraction of the total mass contained in the hylotropic envelope increases, the stellar configuration becomes progressively more extended, resembling a larger and larger proto-star.

In this work, we extend the analysis of \cite{2020lionel} by accounting for the effects of external pressure confinement both on the stellar structure and, more importantly, on the GR instability criterion itself. To facilitate the numerical calculation, we introduce the dimensionless variables:
\begin{align}
x &= \alpha r,
&
\rho &= \rho_c\theta^3,
&
P &= P_c\psi,
&
\phi &= \frac{\alpha^3M_r}{4\pi\rho_c},
\label{eq:hylotrope_variables}
\\
\alpha^2
&=
\frac{\pi G\rho_c^2}{P_c},
\label{eq:hylotrope_alpha}
\end{align}
where the subscript $c$ denotes quantities evaluated at the stellar centre. In terms of these variables, the equations of hydrostatic equilibrium and mass continuity become:
\begin{align}
\frac{d\psi}{dx}
&=
-4\frac{\phi\theta^3}{x^2},
\label{eq:hylotrope_hse}
\\
\frac{d\phi}{dx}
&=
x^2\theta^3.
\label{eq:hylotrope_mass}
\end{align}
The hylotropic equation of state can then be written as:
\begin{equation}
\psi
=
\begin{cases}
\theta^4,
    & \phi\leq\phi_{\rm core}, \\[3pt]
\left(\dfrac{\phi}{\phi_{\rm core}}\right)^{2/3}\theta^4,
    & \phi>\phi_{\rm core},
\end{cases}
\label{eq:hylotrope_dimensionless_eos}
\end{equation}
where:
\begin{equation}
\phi_{\rm core}
=
\frac{\alpha^3M_{\rm core}}{4\pi\rho_c}
\label{eq:phi_core}
\end{equation}
is the dimensionless core mass. The parameter $\phi_{\rm core}$ labels the sequence of hylotropic configurations, ranging from a fully thermally relaxed $n=3$ polytrope to a star composed of a small relaxed core surrounded by a massive, extended envelope. Due to stability constraints, the physically allowed range is:
\begin{equation}
0.459
\lesssim
\phi_{\rm core}
\lesssim
2.02,
\label{eq:phi_core_range}
\end{equation}
where the lower bound corresponds to the minimum core mass required for the configuration to remain gravitationally bound, while the upper bound is reached in the limit of a fully thermally relaxed $n=3$ polytrope.
The corresponding core-mass fraction is:
\begin{equation}
f_{\rm core}
=
\frac{M_{\rm core}}{M_*}
=
\frac{\phi_{\rm core}}{\phi_s},
\label{eq:core_fraction}
\end{equation}
where $\phi_s$ is the dimensionless mass enclosed within the stellar surface. We note that the radius of the star scales rapidly with the core fraction, reflecting the size of the bloated envelope. From our numerical calculations, we find roughly:
\begin{align}
    R_* \propto f_{\rm core}^{-2} \qquad {\rm for} \, \, f_{\rm core}\lesssim0.1.
\end{align}
At the centre of the star, the density and pressure are maximal while the enclosed mass vanishes:
\begin{equation}
\theta(0)=1,
\qquad
\psi(0)=1,
\qquad
\phi(0)=0.
\label{eq:hylotrope_central_bc}
\end{equation}
For an isolated star, the surface is defined by vanishing pressure $\psi(x_s)=0$. Instead, a
pressure-confined hylotrope terminates at:
\begin{equation}
\psi(x_s)
=
q_{\rm ext},
\qquad
q_{\rm ext}
\equiv
\frac{P_{\rm ext}}{P_c}.
\label{eq:hylotrope_pressure_bc}
\end{equation}
Importantly, the family of hylotropic solutions labelled by $\phi_{\rm core}$ describe only the dimensionless stellar structure. To recover a unique physical scale we need to fix the central pressure and impose the marginal-stability condition to determine the stellar mass at which the GR instability is triggered.

\subsection{GR instability of pressure-confined hylotropes}
\label{S:hylotropic_instability}

The general stability condition derived in Section~\ref{S:gr_stability} can be expressed in dimensionless form and simplified further by adopting the dimensionless hylotropic variables. The full derivation is presented in Appendix~\ref{S:derivations}. Here, we state only the form used in the numerical calculations. Following Haemmerl\'e, we first define the dimensionless structure integrals:
\begin{align}
J_1
&\equiv
\int_0^{x_s}
\frac{\beta}{\beta_c}
\psi x^2\,dx,
\label{eq:hylotrope_J1}
\\
J_2
&\equiv
\int_0^{x_s}
x\psi\phi\,dx,
\label{eq:hylotrope_J2}
\\
J_3
&\equiv
\int_0^{x_s}
\phi^2\theta^3\,dx.
\label{eq:hylotrope_J3}
\end{align}
These integrals are evaluated over the pressure-confined configuration, from the stellar centre to the outer boundary at $x=x_s$. We also introduce the post-Newtonian expansion parameter:
\begin{equation}
\sigma
\equiv
\frac{P_c}{\rho_c c^2}.
\label{eq:hylotrope_sigma}
\end{equation}
As shown in Appendix~\ref{S:derivations}, the marginal-stability condition then reduces to:
\begin{align}
0
\approx{}&
\beta_cJ_1
-
\frac{32}{3}\sigma
\left(
J_2+3J_3
\right)
+
\frac{2}{3}(4-s)
q_{\rm ext}x_s^3,
\label{eq:general_marginal_stability}
\end{align}
at leading order in $\beta_c$, $\sigma$ and $q_{\rm ext}$.
The first term represents stabilisation by gas pressure, the second is the leading relativistic destabilisation, and the final term describes the effect of external pressure confinement.

\subsection{Physical scale of marginally stable hylotropes}
\label{S:numerical_method}
\begin{figure}
    \centering
    \includegraphics[width=1\columnwidth]{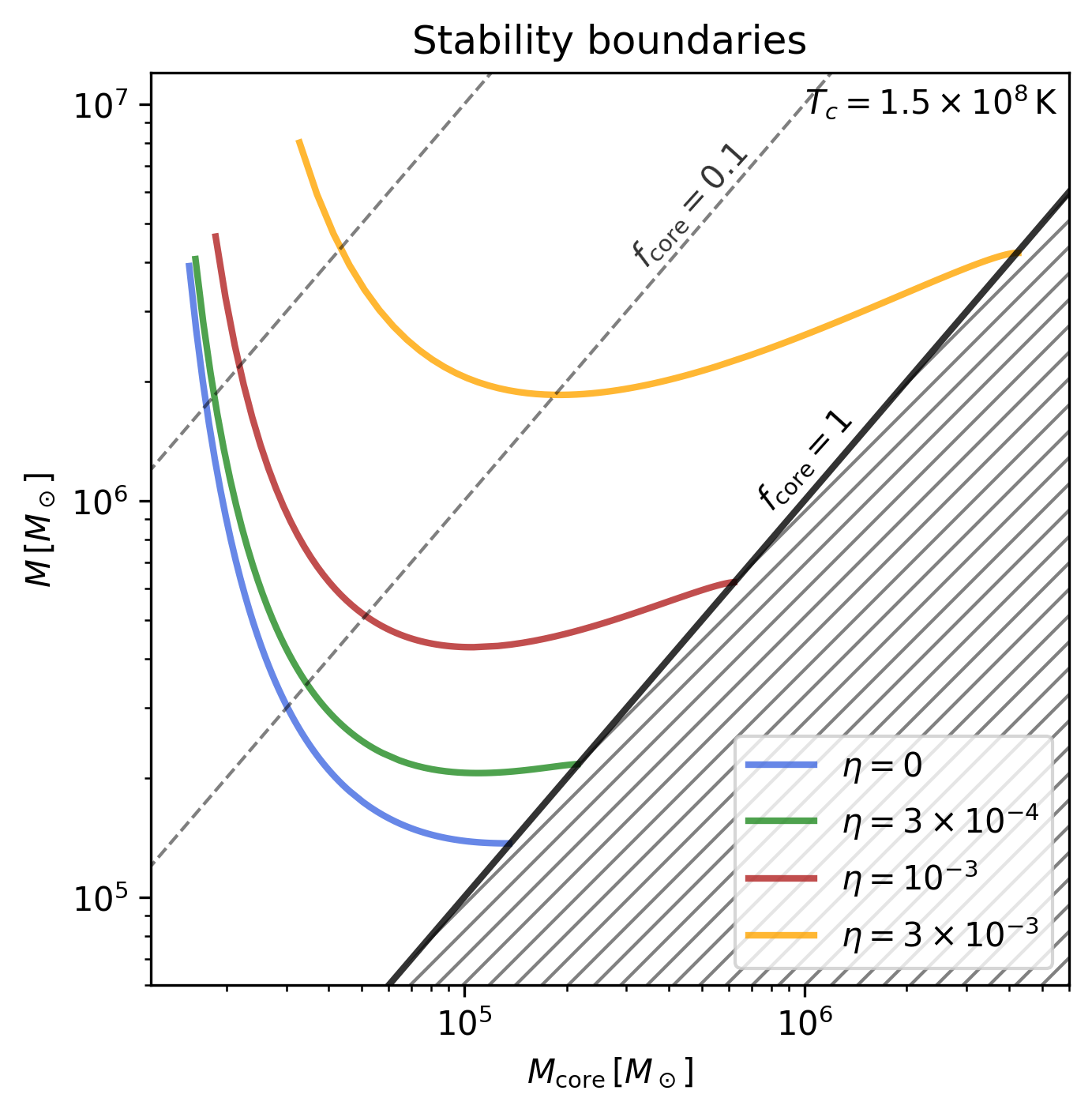}
    \caption{Stability boundaries (coloured lines) for hylotropic SMSs in the core-mass--total-mass plane. For a given value of the ratio between the external and internal pressure, $\eta=P_{\rm ext}/P_*$, stars lying above and to the right of the corresponding boundary are disallowed by the GR instability. Increasing $\eta$ shifts the stability boundary toward larger masses, illustrating the stabilising effect of external pressure confinement (provided that $4-s > 0$, here for $s=0$). The black solid line denotes the fully relaxed $n=3$ polytropic limit with $f_{\rm core}=1$. The stability boundaries terminate in the top left quadrant when the hylotropic models become gravitationally unbound at $\phi_{\rm core}\simeq0.459$.}
    \label{fig:stability_boundary}
\end{figure}

\begin{figure}
    \centering
    \includegraphics[width=1\columnwidth]{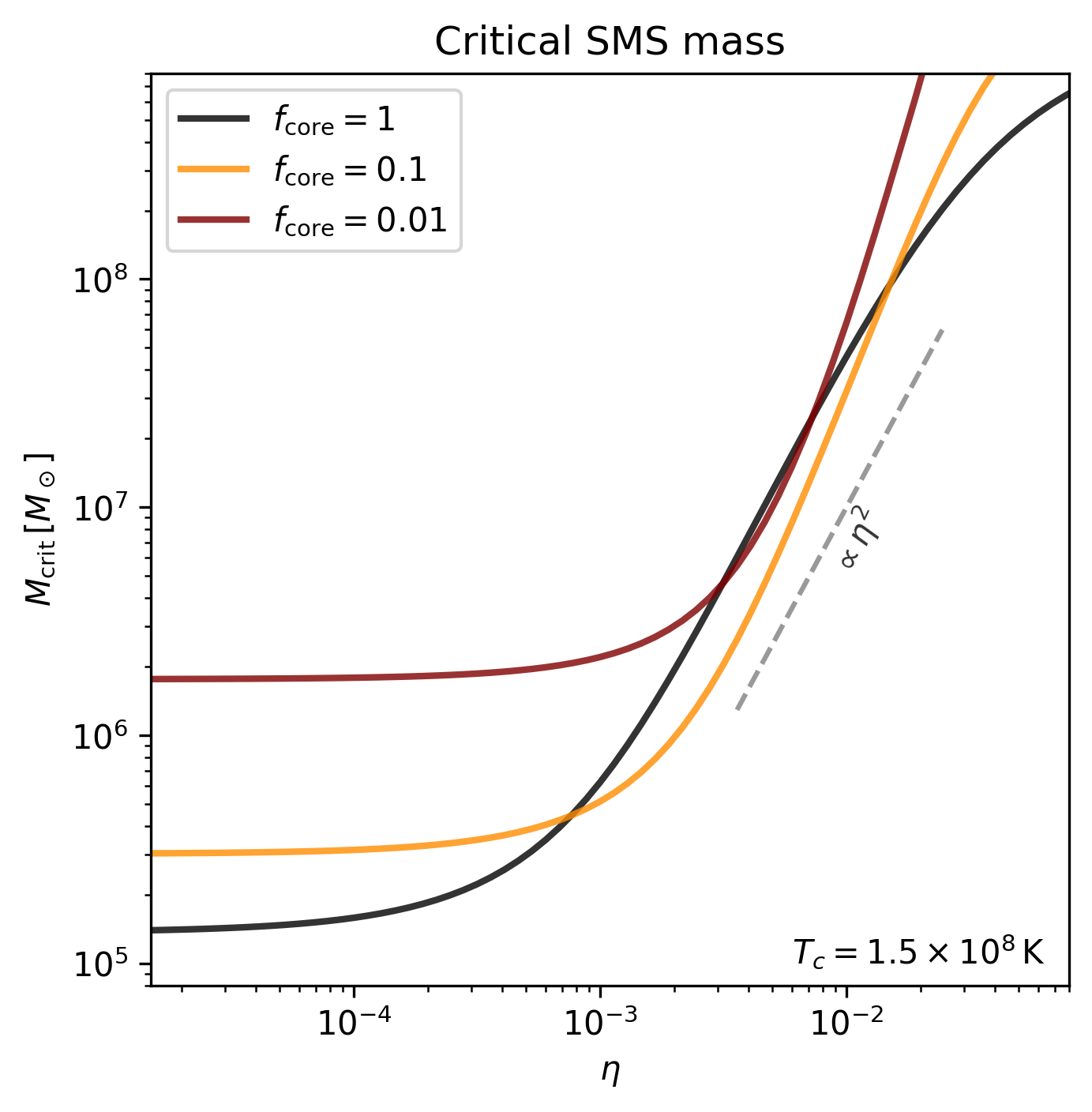}
    \caption{Marginal mass of hylotropic SMSs (solid lines) for several fixed core-mass fractions, shown as a function of the pressure ratio $\eta$. For $4-s>0$ (here shown for $s=0$), pressure confinement delays the onset of the GR instability once $\eta\gtrsim10^{-3}$, allowing the star to reach substantially larger masses before collapse. For the fully relaxed polytropic case, the steep increase in the critical mass follows $M_{\rm crit}\propto\eta^2$.}
    \label{fig:crit_mass}
\end{figure}

We now detail how to recover the physical mass and radius of marginal hylotropes at the verge of the GR instability. The first step is to specify a central temperature $T_c$ and a mean molecular weight $\mu$. Throughout this work, we adopt the standard values of:
\begin{equation}
T_c
=
1.5\times10^8\,{\rm K},
\qquad
\mu
=
0.60,
\label{eq:fiducial_thermodynamic_parameters}
\end{equation}
representative pop III stellar evolution \citep{begelman2010,2020lionel} for helium burning cores. They provide an appropriate scale for large SMS ($\sim 10^6$ M$_{\odot}$) forming and accreting in a high-redshift halo before significant metal enrichment. We will discuss how the results scale with temperature specifically in section \ref{sec:temp}. For fixed $T_c$ and $\mu$, the Eddington and post-Newtonian expansion parameters are related by:
\begin{equation}
\beta_c\sigma
=
Q_T,
\qquad
Q_T
\equiv
\frac{k_{\rm B}T_c}{\mu m_{\rm H}c^2}.
\label{eq:thermal_closure}
\end{equation}
Together with the marginal-stability condition, these relations fix $\beta_c$ and $\sigma$ for each value of $\phi_{\rm core}$, and therefore sets the physical scale of the corresponding hylotrope. 

For each choice of $\phi_{\rm core}$ and $q_{\rm ext}$, integrating the equation of hydrostatic equilibrium determines all the relevant quantities:
\begin{equation}
x_s,\qquad
\phi_s,\qquad
J_1,\qquad
J_2,\qquad
J_3.
\end{equation}
Then, we can substitute $\beta_c=Q_T/\sigma$ into
Eq.~\eqref{eq:general_marginal_stability} to obtain a quadratic equation to determine marginal stability:
\begin{equation}
A\sigma^2
-
C\sigma
-
Q_TJ_1
=
0,
\label{eq:sigma_quadratic}
\end{equation}
where:
\begin{align}
A
&=
\frac{32}{3}
\left(
J_2+3J_3
\right),
\label{eq:quadratic_A}
\\
C
&=
\frac{2}{3}(4-s)
q_{\rm ext}x_s^3.
\label{eq:quadratic_C}
\end{align}
Once $\sigma$ and $\beta_c$ have been determined by taking the positive root of Eq. \eqref{eq:sigma_quadratic}, the dimensional central quantities follow from:
\begin{equation}
P_c
=
\frac{a_{\rm rad}T_c^4}{3(1-\beta_c)},
\qquad
\rho_c
=
\frac{P_c}{\sigma c^2},
\label{eq:central_dimensional_quantities}
\end{equation}
from which $M_{\rm core}$, $M_*$, and $R_*$ are obtained through Eqs.~\eqref{eq:hylotrope_variables} and \eqref{eq:hylotrope_alpha}. To determine the stability boundaries of each family of hylotropes, we repeat this procedure as a function of $\phi_{\rm core}$ for a grid of external-pressure values $q_{\rm ext}$. For every choice of core-mass fraction and external pressure, this completely specifies the hydrostatic SMS configuration at the onset of collapse. The result is a family of marginally stable SMS solutions parametrised by $\phi_{\rm core}$ for each value of $q_{\rm ext}$.

Once the marginal solutions have been recovered, we can distinguish the central pressure $P_c$ from the characteristic hydrostatic pressure scale $P_*$ introduced in the heuristic derivation in Section~\ref{S:GR_Instability:heuristic}. From hydrostatic equilibrium, we define the pressure scale:
\begin{equation}
P_*
=
\alpha_W
\frac{GM_*^2}{4\pi R_*^4},
\qquad
\alpha_W=\frac{3}{2},
\label{eq:hydrostatic_pressure_scale}
\end{equation}
which relates the characteristic stellar pressure to the gravitational energy density. The coefficient $\alpha_W=3/2$ is chosen to reproduce the gravitational binding energy of an $n=3$ polytrope. This definition is therefore exact in the limit $f_{\rm core}=1$, while for more extended hylotropic configurations it should be regarded as a fiducial global pressure scale. To compare this scale directly with the external pressure confinement we define the dimensionless pressure ratio:
\begin{equation}
\eta
\equiv
\frac{P_{\rm ext}}{P_*}
=
q_{\rm ext}\frac{P_c}{P_*}.
\label{eq:pressure_confinement_eta}
\end{equation}
As anticipated from the heuristic argument, we expect external confinement to produce an appreciable shift in the stability boundary once $\eta$ approaches approximately $10^{-3}$.

The numerical implementation in \texttt{Python} for these calculations is straightforward. For each choice of $\phi_{\rm core}$ and $q_{\rm ext}$, we integrate the dimensionless equations of hydrostatic equilibrium and mass continuity until $\psi(x_s)=q_{\rm ext}$. We then evaluate the structure integrals $J_1$, $J_2$, and $J_3$, solve Eq.~\eqref{eq:sigma_quadratic} for $\sigma$, and use Eqs.~\eqref{eq:central_dimensional_quantities} to recover the corresponding physical mass, radius, and central quantities. The results are discussed in the following section.

\subsection{Stability boundaries and critical mass}

Fig.~\ref{fig:stability_boundary} shows the stability boundaries for hylotropic SMSs. The boundary is parametrized by the core mass fraction, tracing a curve in the core mass - total mass plane. Accreting SMS would evolve through this plane from the lower left toward the upper right, with a characteristic core-mass fraction set by the thermal state of the star. Collapse then occurs once the evolutionary track crosses the corresponding stability boundary. In the plot, we show the stability boundary for $\eta=0$, recovering exactly the previous results of \cite{2020lionel}. We also show several boundaries that account for the new modified stability criterion accounting for pressure confinement, here for $s=0$. As the pressure ratio $\eta$ increases, the stability boundary shifts toward larger masses, and we can indeed notice order-unity modification from the vacuum result for $\eta\sim10^{-3}$, in agreement with the heuristic estimate. These results show how external pressure confinement stabilises stars, allows to reach a larger mass before collapse via the GR instability.

We further quantify the effect of the new stability criterion in Fig.~\ref{fig:crit_mass}, where we show the critical mass as a function of $\eta$ for several fixed values of $f_{\rm core}$ and $s=0$. For a fully relaxed polytrope, $f_{\rm core}=1$, a clear break appears at $\eta\sim10^{-3}$. Beyond this characteristic pressure scale, the critical mass rises rapidly, approximately as $M_{\rm crit}\propto\eta^2$, reaching several $10^8\,{\rm M_\odot}$ before eventually flattening. Similar behaviour is also visible for the other core-mass fractions. The figure shows how, once $\eta\gtrsim10^{-3}$, pressure confinement competes directly with the destabilisation due to relativistic effects, which can lead to substantially larger masses at marginal stability.

The approximate $\eta^2$ scaling for a pure polytrope can be derived directly from the stability criterion in Eq.~\eqref{eq:general_marginal_stability}, by comparing the GR destabilisation term and the external pressure confinement term. Firstly, recall that:
\begin{equation}
q_{\rm ext}
=
\frac{P_s}{P_c}
=
\eta\frac{P_*}{P_c},
\end{equation}
which simply states that the confinement term is proportional to $\eta$. The stability condition can therefore be written schematically as:
\begin{equation}
\beta_c
\simeq
A\sigma-B\eta,
\label{eq:heuristic_eta_stability}
\end{equation}
where $A$ and $B$ depend only on the chosen hylotropic structure and on the pressure-response index $s$. When the pressure is relevant, marginal stability occurs when the GR correction matches the confinement term:
\begin{equation}
A\sigma
\sim
B\eta.
\end{equation}
Additionally, at fixed central temperature and composition, Eq.~\eqref{eq:thermal_closure} gives
\begin{equation}
\beta_c\sigma
=
Q_T
=
{\rm const},
\end{equation}
so that:
\begin{equation}
\sigma\propto\eta,
\qquad
\beta_c
=
\frac{Q_T}{\sigma}
\propto\eta^{-1}.
\end{equation}
Using Eqs.~\eqref{eq:hylotrope_alpha} for a fixed dimensionless structure gives:
\begin{equation}
M_*
\propto
\frac{P_c^{3/2}}{\rho_c^2},
\end{equation}
and with $\rho_c=P_c/(\sigma c^2)$ from Eq.~\eqref{eq:central_dimensional_quantities} this becomes:
\begin{equation}
M_*
\propto
\sigma^2P_c^{-1/2}.
\end{equation}
At fixed $T_c$, the central pressure varies only weakly in the radiation-dominated limit $\beta_c\ll1$:
\begin{equation}
P_c
=
\frac{a_{\rm rad}T_c^4}{3(1-\beta_c)} \approx \frac{a_{\rm rad}T_c^4}{3}.
\end{equation}
Hence:
\begin{equation}
M_{\rm crit}
\propto
\sigma^2
\propto
\eta^2,
\label{eq:critical_mass_eta_scaling}
\end{equation}
and we recover the scaling seen in Fig. \ref{fig:crit_mass}.

\subsection{Connection to high-redshift accretion flows}
\label{S:highz_accretion}
\begin{figure}
    \centering
    \includegraphics[width=1\columnwidth]{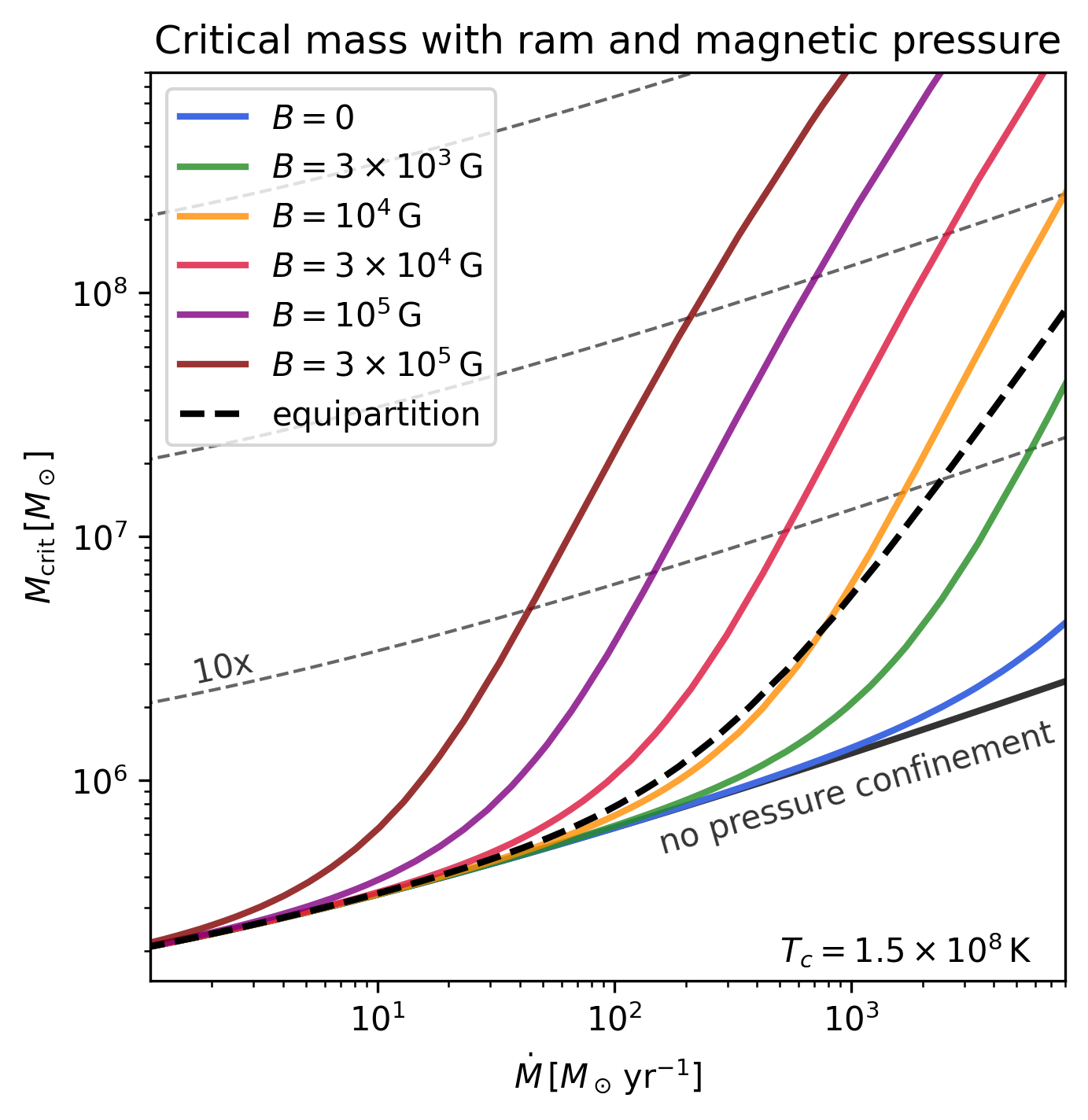}
    \caption{Marginal mass of hylotropic SMSs confined by ram pressure and an isotropic magnetic pressure (colored lines), assuming the $\dot M$--$f_{\rm core}$ relation inferred from the \texttt{GENEC} models (see text for details). The grey solid line indicates the corresponding results without accounting for the modified instability criterion derived in this work. The thin dashed grey lines indicate multiples of the latter results. Additionally, the thick dashed black line shows the marginal mass assuming equipartition of magnetic and thermal energy for an $\alpha$-disk, assuming $\alpha=0.1$ and a scale height of $h=0.1$. More details are provided in the text.}
    \label{fig:critmass_mdotB}
\end{figure}
In this section, we attempt to tie our results to realistic high-redshift accretion environments. The formation of SMSs requires sustained gas inflows of order $\gtrsim 0.1\,{\rm M_\odot}\,{\rm yr}^{-1}$, which may arise, for example, in the cores of atomic cooling haloes \citep{2013prieto,2020regan,2022latif,2024prole} exposed to a sufficiently intense Lyman-Werner radiation~\citep{Omukai2001,OhHaiman2002,BrommLoeb:2003,Jemma2011}
or during gas-rich galaxy mergers \citep{Mayer2010,2014bonoli,Mayer2015,2023zwick,2024ApJ...961...76M}. To model this scenario consistently we consider two main sources of external pressure confinement: the ram pressure exerted by the inflowing gas and an approximately isotropic magnetic pressure associated with magnetic fields advected or generated by the accretion flow. For approximately spherical accretion, the ram pressure is \citep{Bondi:1952wx}:
\begin{equation}
P_{\rm ram}
=
\frac{\dot M}{4\pi R_*^2}
\left(
\frac{2GM_*}{R_*}
\right)^{1/2},
\label{eq:ram_pressure}
\end{equation}
where we have assumed that the inflow velocity at the stellar surface is close to free fall. Since $P_{\rm ram}\propto R_*^{-5/2}$ at fixed $M_*$ and $\dot M$, the corresponding index is $s=5/2$. Ram pressure therefore provides a stabilizing contribution. Following \citet{Pringle:BLs:1989} and \citet{Armitage:BLs:2002}, we consider a simple model for a magnetized accretion disk depositing material onto a star that is rotating slower than the Keplerian velocity. The key aspect is that gas deposited onto the star must then pass through a narrow boundary layer with positive shear. Then, differential rotation in this layer can wind and amplify the magnetic field into predominantly toroidal configurations to produce a confining magnetic pressure. For magnetic pressure:
\begin{equation}
P_B
=
\frac{B^2}{8\pi}.
\label{eq:magnetic_pressure}
\end{equation}
As a simple limiting case, we assume that this pressure remains approximately fixed during the radial perturbation, corresponding to $s\simeq0$, which is also a stabilizing contribution.

As discussed in Section~ \ref{S:GR_Instability:heuristic} and \ref{S:hylotropic_instability}, external confinement begins to appreciably modify the GR instability once:
\begin{equation}
\eta
\equiv
\frac{P_{s}}{P_*}
\sim10^{-3}.
\label{eq:relevant_pressure_ratio}
\end{equation}
Dividing by the stellar scale pressure, ram pressure can be written as:
\begin{equation}
\frac{P_{\rm ram}}{P_*}
=
\frac{\sqrt{2}}{\alpha_W}
\frac{\dot M R_*^{3/2}}
{G^{1/2}M_*^{3/2}},
\label{eq:ram_pressure_ratio}
\end{equation}
while for magnetic pressure:
\begin{equation}
\frac{P_B}{P_*}
=
\frac{B^2R_*^4}
{2\alpha_WGM_*^2}.
\label{eq:magnetic_pressure_ratio}
\end{equation}
We see in both cases that the importance of pressure confinement increases very rapidly with stellar radius. For reference, we scale the result to a marginally stable hylotrope with $f_{\rm core}=0.1$. From our numerical calculations, such a configuration has approximately:
\begin{equation}
M_*\simeq3\times10^5\,{\rm M_\odot},
\qquad
R_*\simeq20\,{\rm AU}.
\end{equation}
The ram-pressure contribution then becomes:
\begin{align}
\frac{P_{\rm ram}}{P_*}
\simeq{}&
10^{-3}
\left(
\frac{\dot M}{1.2\times10^4\,{\rm M_\odot}\,{\rm yr}^{-1}}
\right)
\nonumber\\
&\times
\left(
\frac{M_*}{3\times10^5\,{\rm M_\odot}}
\right)^{-3/2}
\left(
\frac{R_*}{20\,{\rm AU}}
\right)^{3/2},
\label{eq:ram_pressure_ratio_scaled}
\end{align}
while the magnetic contribution is:
\begin{align}
\frac{P_B}{P_*}
\simeq{}&
10^{-3}
\left(
\frac{B}{9.4\times10^4\,{\rm G}}
\right)^2
\left(
\frac{M_*}{3\times10^5\,{\rm M_\odot}}
\right)^{-2}
\nonumber\\
&\times
\left(
\frac{R_*}{20\,{\rm AU}}
\right)^4.
\label{eq:magnetic_pressure_ratio_scaled}
\end{align}
For these fiducial values, accretion rates of order $10^4\,{\rm M_\odot}\,{\rm yr}^{-1}$ or magnetic fields of order $10^5\,{\rm G}$ are required to reach the regime in which pressure confinement significantly affects the GR instability. Recall, however, that both the marginal mass and the stellar radius depend strongly on the core-mass fraction. In particular, the radius scales approximately as $R_*\propto f_{\rm core}^{-2}$ for the extended hylotropes of interest.

For the purposes of this work, we adopt a simple strategy to eliminate the additional dependence of the results on the core mass fraction. Several works have already investigated the evolution of rapidly accreting SMSs numerically \citep{hosokawa2013,haemmerle2018a,2023herrington,2024A&A...689A.351N,2025arXiv251108516N}, and in particular, \cite{2020lionel} displays the results of calculations performed with the \texttt{GENEC} \citep{2008Ap&SS.316...43E} stellar evolution code: At fixed accretion rate, SMS evolve with an approximately constant core-mass fraction, only slightly increasing as the mass of the star grows. Following these results, it is possible to approximately associate a given accretion rate with a characteristic core mass fraction for hylotropes close to the stability boundary. We therefore simply interpolate the results of \texttt{GENEC} models over the range of accretion rates shown by \cite{2020lionel}, finding the following approximate relation:
\begin{equation}
f_{\rm core}
\simeq
0.2
\left(
\frac{\dot M}{{\rm M_\odot}\,{\rm yr}^{-1}}
\right)^{-0.38},
\label{eq:fcore_mdot_fit}
\end{equation}
interpolated for:
\begin{equation}
0.1
\lesssim
\frac{\dot M}{{\rm M_\odot}\,{\rm yr}^{-1}}
\lesssim
10^3.
\end{equation}
We assume that this scaling provides a representative mapping between accretion rate and hylotropic structure close to marginal stability. Note the same procedure could be repeated by interpolating results from a different stellar-evolution code. Ultimately, the most consistent approach would be to supply a stellar evolution code with the modified stability criterion derived in this work, an endeavor that is planned for future work. For the purposes of this work, however, Eq.~\eqref{eq:fcore_mdot_fit} is sufficient to close the problem and eliminate the additional dependence on $f_{\rm core}$. With this simplification, the two external parameters, $\dot M$ and $B$, fully determine the stellar mass at the onset of collapse.

The resulting marginal masses are shown in Fig.~\ref{fig:critmass_mdotB} as a function of accretion rate for several values of the magnetic field at the stellar surface. Overall, the marginal mass increases with accretion rate. The effect of the modified stability criterion appears as a characteristic break in the slope of each curve, beyond which the marginal mass rises much more rapidly. In the absence of magnetic fields, i.e. for ram pressure alone, this transition occurs at accretion rates of several $10^3\,{\rm M_\odot}\,{\rm yr}^{-1}$. Increasing the magnetic field shifts the onset of this rapid rise to lower accretion rates, so that the final marginal mass is determined jointly by $\dot M$ and $B$. For example, increasing the marginal mass by a factor of $\sim10$ relative to the case without pressure confinement requires surface magnetic fields of order $3\times10^5\,{\rm G}$ at $\dot M\simeq30\,{\rm M_\odot}\,{\rm yr}^{-1}$, and of order $3\times10^4\,{\rm G}$ at $\dot M\simeq300\,{\rm M_\odot}\,{\rm yr}^{-1}$. The numerical results are well described, to within a few percent, by a smoothly broken power law:
\begin{align}
&M_{\rm crit}(\dot M,B)
=
\nonumber \\&A
\left(
\frac{\dot M}{{\rm M_\odot}\,{\rm yr}^{-1}}
\right)^a
\left[
1+
\left(
\frac{\dot M}{\dot M_{\rm turn}(B)}
\right)^q
\right]^{\Delta a/q},
\label{eq:final_mass_fit}
\end{align}
with:
\begin{equation}
A
=
1.9\times10^5\,{\rm M_\odot},
\quad
a=0.3,
\quad
\Delta a=1.4,
\quad
q=2.0.
\end{equation}
The turnover accretion rate depends on the magnetic field according to:
\begin{align}
&\dot M_{\rm turn}(B)
=\nonumber \\
&7.6\times10^3\,{\rm M_\odot}\,{\rm yr}^{-1}
\left[
1+
\left(
\frac{B}{5.0\times10^2\,{\rm G}}
\right)^{1.3}
\right]^{-1/1.3}.
\label{eq:turnover_mdot_fit}
\end{align}

Finally, we further reduce the number of independent external parameters by introducing a simple relation between the magnetic field strength and the accretion flow that supplies the mass accretion rate. As shown in \citet{Takasao:BLs:2025} the amplification of the magnetic field saturates once the magnetic pressure becomes comparable to the thermal pressure in the disk. For concreteness, we treat the accretion flow as a standard $\alpha$-disk. Then, for a scale height $H\equiv hR_{\rm d}$, the thermal pressure near the stellar surface is approximately \citep{Shakura:1973uy}:
\begin{align}
    P \simeq
    \frac{\dot M}{6\pi\alpha h}
    \sqrt{\frac{GM_*}{R_*^5}} \, ,
\end{align}
where $\alpha$ characterizes the efficiency of angular-momentum transport. Equating the magnetic and thermal pressures of this simple disk model then gives a characteristic equipartition field strength:
\begin{align}
    B_{\rm eq}
    &\simeq
    7.8\times10^4\,
    \left(\frac{\dot M}{100\,{\rm M_\odot}\,{\rm yr}^{-1}}\right)^{1/2}
    \left(\frac{M_\ast}{10^6\,{\rm M_\odot}}\right)^{1/4}
    \nonumber\\
    &\qquad\times
    \left(\frac{R_\ast}{20\,{\rm AU}}\right)^{-5/4}
    \left(\frac{\alpha h}{0.01}\right)^{-1/2}
    \,{\rm G}\, ,
    \label{eq:Beq}
\end{align}
where we adopt the fiducial values $\alpha=0.1$ and $h=0.1$. Comparison with Fig.~\ref{fig:critmass_mdotB} shows that fields near equipartition can reach strengths large enough for magnetic pressure confinement to become important at accretion rates of order $\dot M\sim100\,{\rm M_\odot}\,{\rm yr}^{-1}$. Note however, that in this estimate we are extrapolating from a disk-like accretion flow to the approximately isotropic and spherically symmetric pressure boundary assumed in our stellar stability calculation. Nevertheless, at the level of energetics alone, the required magnetic pressure is available in sufficiently extreme but plausible accretion flows, as discussed in Section~\ref{S:discussion}. In Fig.~\ref{fig:critmass_mdotB}, we show the critical mass obtained under the equipartition assumption. For our fiducial choices of $\alpha$ and $h$, pressure confinement begins to appreciably delay the onset of the GR instability at accretion rates $\dot M\gtrsim100\,{\rm M_\odot}\,{\rm yr}^{-1}$, while an increase in the critical mass by a factor of $\sim10$ is reached at accretion rates of a few $10^3\,{\rm M_\odot}\,{\rm yr}^{-1}$.

\subsection{Scaling with central temperature}
\label{sec:temp}
In this work we have fixed the central temperature and presented results based on the characteristic scale of helium burning cores. Here, we present a scaling argument to estimate how the central temperature affects the mass at marginal stability, considering the limiting cases of negligible and dominant pressure confinement. We start from Eq.~\eqref{eq:sigma_quadratic}, which determines the post-Newtonian parameter at marginal stability:
\begin{equation}
A\sigma^2
-
C\sigma
-
Q_TJ_1
=
0,
\end{equation}
where $Q_T\propto T_c$ from Eq.~\eqref{eq:thermal_closure}. Taking the positive root gives:
\begin{equation}
\sigma
=
\frac{
C+\sqrt{C^2+4AJ_1Q_T}
}{
2A
}.
\label{eq:sigma_temperature_scaling}
\end{equation}
At the same time, Eqs.~\eqref{eq:hylotrope_alpha} and \eqref{eq:central_dimensional_quantities} imply, for a fixed dimensionless stellar structure:
\begin{equation}
M_*
\propto
\sigma^2P_c^{-1/2}.
\end{equation}
In the radiation-pressure-dominated limit, $\beta_c\ll1$, the central pressure scales as $P_c\propto T_c^4$, and therefore:
\begin{equation}
M_{\rm crit}
\propto
\frac{\sigma^2}{T_c^2}.
\label{eq:mass_temperature_general}
\end{equation}
In the limit of negligible pressure confinement, $C^2\ll4AJ_1Q_T$, Eq.~\eqref{eq:sigma_temperature_scaling} gives:
\begin{equation}
\sigma
\propto
Q_T^{1/2}
\propto
T_c^{1/2},
\end{equation}
and hence:
\begin{equation}
M_{\rm crit}
\propto
T_c^{-1}.
\label{eq:mass_temperature_vacuum}
\end{equation}
Conversely, when pressure confinement dominates, $C^2\gg4AJ_1Q_T$, the leading contribution to Eq.~\eqref{eq:sigma_temperature_scaling} is $\sigma\simeq C/A$, which is independent of $T_c$. In this limit:
\begin{equation}
M_{\rm crit}
\propto
T_c^{-2}.
\label{eq:mass_temperature_confined}
\end{equation}
We therefore find two limiting regimes. Lower central temperatures generally allow the SMS to reach larger masses before the onset of the GR instability, with approximately $M_{\rm crit}\propto T_c^{-1}$ in the vacuum limit and a stronger $M_{\rm crit}\propto T_c^{-2}$ dependence when pressure confinement dominates. The central temperature therefore introduces an important additional dependence of the final collapse mass. For realistic SMS, nuclear burning constrains the central temperature to a characteristic scale of order $10^8\,{\rm K}$, with our fiducial value $T_c=1.5\times10^8\,{\rm K}$ being representative of a helium-burning core \citep{2020lionel}. We therefore expect our results to capture the relevant order of magnitude and, more importantly, the effect of pressure confinement. The scalings found here can also be applied to the fitting formula to adapt them to different central temperatures. Nevertheless, for precise quantitative results $T_c$ should ultimately be determined self-consistently with a stellar-evolution calculation, as its value depends on the evolutionary state and accretion history of the SMS.

\section{Discussion and Conclusion}
\label{S:discussion}
\subsection{Summary}
The main result of this work is the derivation of a general correction to the GR instability criterion that accounts for pressure confinement.  When applied to hylotropic models of rapidly accreting SMSs, we find that pressure confinement begins to appreciably modify the stability threshold once the external pressure reaches approximately $0.1\%$ of the characteristic stellar pressure. We then consider the confinement caused by the ram pressure and magnetic pressure associated to the accretion flow responsible for assembling the SMS. As a rough estimate, we find that delaying the GR instability through pressure confinement generally requires combined accretion rates and magnetic field strengths of order $\dot M\gtrsim100\,{\rm M_\odot}\,{\rm yr}^{-1}$ and $B\gtrsim10^4\,{\rm G}$.

\subsection{Conditions in high-z accretion flows}
Here we argue that such conditions may plausibly arise in a rare subset of high-redshift halos. Regarding the accretion rate, it is known that major mergers of gas-rich galaxies can trigger exceptionally strong inflows toward galactic centers, with the most extreme estimates reaching thousands of ${\rm M_\odot}\,{\rm yr}^{-1}$ \citep{2010hopkins,Mayer2010,Mayer2015,2021MNRAS.508.3672P,lionel2021}. In addition, cold cosmological accretion can supply massive high-redshift galaxies with baryonic inflow rates of hundreds of ${\rm M_\odot}\,{\rm yr}^{-1}$ \citep{2012ApJ...745L..29D,2014MNRAS.440.1865F,2025MNRAS.537.2726W}. In this latter case, the remaining requirement is an efficient mechanism capable of transporting a significant fraction of this gas from galactic scales down to the vicinity of the SMS. However, the existence of massive high-redshift quasars directly implies that such efficient angular-momentum transport must operate in at least a subset of galaxies, allowing gas to reach the central object at sustained near-Eddington rates.

Likewise, strong magnetic fields are expected in such rapid inflow scenarios, as initially weak galactic-scale fields can be advected inward and strongly amplified by the flow \citep{2011arXiv1102.3558S,2024OJAp....7E..18H,2024OJAp....7E..19H,2024ApJ...973..141G}. In particular, \citet{2024OJAp....7E..19H} find inflow rates of tens of ${\rm M_\odot}\,{\rm yr}^{-1}$ together with magnetic field strengths reaching $\sim10^3\,{\rm G}$ on scales of hundreds of AU. These values are consistent with our equipartition argument, which predicts that accretion rates of order $\dot M\sim10\,{\rm M_\odot}\,{\rm yr}^{-1}$ correspond to magnetic field strengths of several $10^3\,{\rm G}$ when the magnetic pressure approaches the thermal pressure of the inner accretion flow. Note also that these simulations were designed primarily to study typical accretion onto quasars, rather than SMS-forming environments. Nevertheless, we stress that these considerations are not intended to claim that the conditions required to delay SMS collapse are typical. Rather, they identify the circumstances under which a rare fraction of the halo population may produce correspondingly extreme central objects. The regime of greatest interest is therefore precisely the tail of the distribution.

\subsection{Caveats}
In this work, the results in Section~\ref{S:GR_Instability} and the majority of Section~\ref{S:hylotropes} are mathematically rigorous within the stated assumptions. Instead, the  estimates in Section~\ref{S:highz_accretion} rely on an interpolation between the accretion rate and the core-mass fraction based on results displayed in \cite{2020lionel}. Additionally, all results are computed for a fixed central temperature. While these simplifications provide a useful way to connect the modified stability criterion to rapidly accreting SMSs, the detailed mapping between $\dot M$ and the stellar structure is model dependent and ultimately should be determined in a slef consistent stellar evolution calculation. In addition, both the hylotropic models and the pressure confinement calculation assume spherical symmetry and homologous displacements, whereas realistic SMSs are expected to rotate and to accrete through strongly anisotropic, magnetized flows. Similarly, our equipartition argument also relies on a disc geometry, and is meant only as a crude estimate. Therefore, an important next step is to implement the modified boundary condition directly in stellar-evolution calculations that include rotation~\citep{Chandra1969} and a more self-consistent treatment of the accretion flow. Additionally, the validity of the new stability criterion needs to be tested in regimes in which pressure confinement is anisotropic. This would allow pressure confinement, angular-momentum transport, and the evolving entropy structure of the star to be followed simultaneously, rather than incorporated through separate approximate prescriptions. Finally, the pressure boundary calculation should be extended to non-homologous trial displacements. The importance of the latter is shown explicitly in e.g. \citet{2022MNRAS.517.1584N}, where it is found that SMS with structures resulting from evolved stellar models can become GR unstable due to non-homologous pulsations.

\subsection{Implications for BH seeding}
Establishing a self-consistent connection between the properties of the inflow and those of the SMS would make it possible to address the problem at the population level. For example, one could ask what fraction of high-redshift halos experience inflow rates above $\dot M\sim100\,{\rm M_\odot}\,{\rm yr}^{-1}$, how long such phases persist, what magnetic fields and angular-momentum fluxes accompany them, and what distribution of SMS critical masses follows as a result. This would provide a direct route from cosmological halo and merger statistics to the expected mass and luminosity functions of SMSs and their black-hole remnants. We propose a suggestive scenario in which the seeds of the most massive quasars originate from an especially rare subset of SMSs, for which rotational support or pressure confinement substantially delays collapse and allows the star to reach unusually large masses. A broader population of direct-collapse seeds with characteristic masses of order $\sim10^5\,{\rm M_\odot}$ could instead arise from more typical SMSs formed under less extreme conditions. In this picture, direct collapse itself remains uncommon, while the exceptionally massive seeds required for the most extreme quasars occupy an even rarer tail of the same underlying population. Light seeds can instead provide the bulk population of lighter massive BH.

An additional interesting and timely aspect is the proposed SMS and quasi-star interpretations of the LRD observed in \textit{JWST} deep fields \citep{NaiduMatthee:2025,2026ApJ...998..124N,2026ApJ...996...48B,2026ApJ..1002....7Z,2026arXiv260321714R}. A potential difficulty for standard SMS interpretations is that non-rotating models typically encounter the GR instability at masses of $\sim10^{4.5}$--$10^{5.5}\,{\rm M_\odot}$, whereas the luminosities inferred for some LRDs correspond to characteristic stellar masses extending to $\sim10^6$--$10^{7}\,{\rm M_\odot}$. However, these estimates treat the SMS stability problem independently of the extreme accretion flow required to create and sustain the object. Our results, together with the importance of rotation, suggest that consistently accounting the accretion flow will postpone the onset of the GR instability and extend the allowed SMS mass range toward that inferred for the brightest LRDs. A population model linking halo inflow histories to SMS structure would therefore also provide a way to test whether the luminosity distribution of LRDs can be reproduced.

\subsection{Conclusion}

Overall, it is reasonable to expect SMSs to form precisely in environments where a large-scale inflow simultaneously delivers mass, angular momentum, and magnetic flux to the central regions. The natural outcome is therefore a rapidly accreting and rotating object that is additionally confined by ram and magnetic pressure. Crucially, both rotation and pressure confinement delay the onset of the GR instability by providing additional stabilising contributions. The interplay between rotation and magnetic fields is particularly suggestive, as previous work has shown that surface magnetic fields of order $\sim10^4\,{\rm G}$ can strongly regulate the angular momentum of an accreting SMS through magnetic braking \citep{haemmerle2019a}, keeping the stellar rotation subcritical. Remarkably, this is also approximately the field scale at which magnetic pressure begins to modify the GR stability threshold in our models. Therefore, we should expect that rotational support, magnetic braking, and pressure confinement may therefore act together in determining the mass at which collapse ultimately occurs. A number of studies have explored the GR instability as a function of accretion rate, stellar structure, and metallicity, with metallicity in particular generally tending to reduce the mass at marginal stability \citep{haemmerle2018a,2022MNRAS.517.1584N,2025arXiv251108516N}. Our results suggest that such predictions should ultimately be revisited together with the stabilising effects naturally associated to the formation of SMS.

To conclude, the broader message of this work is the following: The same accretion flow responsible for the assembly of the SMS supplies mass, angular momentum and provides confinement via ram pressure and magnetic fields. Accounting for these aspects strongly modify the onset of the GR instability that is ultimately responsible for terminating the growth of highly accreting SMS. Determining how these effects operate together in realistic high-redshift environments is essential for predicting the maximum masses of SMSs and the initial masses of the black holes they produce.

\label{S:Methods}
\section*{Acknowledgments}
L.Z. is supported by the European Union’s Horizon 2024 research and innovation program under the Marie Sklodowska-Curie grant agreement No. 101208914. C.T. is supported by the European Union’s Horizon 2023 research and innovation program under the Marie Sklodowska-Curie grant agreement No. 101148364. The Center of Gravity is a Center of Excellence funded by the Danish National Research Foundation under grant no. DNRF184. L.Z. thanks Jaime Roman-Garza and Ralf Klessen for their useful comments. L.Z also thanks Itai Linial for the delightful thought experiment.

\appendix
\section{Derivations}
\label{S:derivations}

\subsection{Chandrasekhar's variational identity
with a pressure-confined boundary}

We write the static, spherically symmetric metric as:
\begin{equation}
ds^2
=
-e^\nu c^2dt^2
+
e^\lambda dr^2
+
r^2d\Omega^2,
\end{equation}
and begin from Chandrasekhar's equation for linear radial pulsations \citep{1964chandra}:
\begin{align}
\omega^2 e^{\lambda-\nu}(P+\epsilon)\xi
={}&
\frac{4}{r}\frac{dP}{dr}\,\xi
-
e^{-(\lambda+2\nu)/2}
\frac{d}{dr}
\left[
e^{(\lambda+3\nu)/2}
\frac{\gamma P}{r^2}
\frac{d}{dr}
\left(
r^2e^{-\nu/2}\xi
\right)
\right]
\nonumber\\
&
+
\frac{8\pi G}{c^4}
e^\lambda P(P+\epsilon)\xi
-
\frac{1}{P+\epsilon}
\left(
\frac{dP}{dr}
\right)^2\xi.
\label{eq:chandrasekhar_pulsation}
\end{align}
Here $P(r)$ and $\epsilon(r)$ are the equilibrium pressure and total energy-density profiles, $\gamma$ is the first adiabatic exponent, $\omega$ is the radial-mode frequency, $\xi(r)$ is the radial Lagrangian displacement, and $\nu(r)$ and $\lambda(r)$ are the equilibrium metric functions.

Following \cite{1964chandra}, we now construct the variational form of the radial pulsation problem. We define:
\begin{equation}
u(r)
\equiv
r^2e^{-\nu/2}\xi,
\label{eq:u_definition}
\end{equation}
multiply Eq.~\eqref{eq:chandrasekhar_pulsation} by $r^2e^{(\lambda+\nu)/2}\xi$, and integrate over the equilibrium stellar domain $0\leq r\leq R_*$. For an isolated configuration, the equilibrium surface satisfies:
\begin{equation}
P(R_*)=0,
\end{equation}
while regularity at the centre and the physical surface boundary condition require:
\begin{equation}
\xi(0)=0,
\qquad
\Delta P(R_*)=0.
\label{eq:chandrasekhar_boundary_conditions}
\end{equation}
Here $\Delta P$ denotes the Lagrangian pressure perturbation, related to
the Eulerian perturbation $\delta P$ by:
\begin{equation}
\Delta P
\equiv
\delta P+\xi\frac{dP}{dr}.
\label{eq:lagrangian_pressure_definition}
\end{equation}
For radial adiabatic perturbations:
\begin{equation}
\delta P
=
-\xi\frac{dP}{dr}
-
\gamma P
\frac{e^{\nu/2}}{r^2}
\frac{du}{dr},
\label{eq:eulerian_pressure_perturbation}
\end{equation}
and therefore:
\begin{equation}
\Delta P
=
-\gamma P
\frac{e^{\nu/2}}{r^2}
\frac{du}{dr}.
\label{eq:lagrangian_pressure_perturbation}
\end{equation}
For an isolated star, the condition $P(R_*)=0$ therefore ensures that $\Delta P(R_*)=0$ for a regular eigenfunction.

The only term in Eq.~\eqref{eq:chandrasekhar_pulsation} containing a differential operator acting on the displacement is the adiabatic-pressure term. Its contribution to the integrated equation is:
\begin{align}
-\int_0^{R_*}
u\frac{d}{dr}
\left[
e^{(\lambda+3\nu)/2}
\frac{\gamma P}{r^2}
\frac{du}{dr}
\right]dr
={}&
\int_0^{R_*}
e^{(\lambda+3\nu)/2}
\frac{\gamma P}{r^2}
\left(\frac{du}{dr}\right)^2dr
\nonumber\\
&
-
\left[
u
e^{(\lambda+3\nu)/2}
\frac{\gamma P}{r^2}
\frac{du}{dr}
\right]_0^{R_*}.
\label{eq:adiabatic_operator_integration_by_parts}
\end{align}
The central contribution vanishes by regularity, while the surface contribution vanishes under vacuum boundary conditions. The resulting variational identity is:
\begin{equation}
\omega^2\mathcal I_0
=
\mathcal S_1+\mathcal S_2+\mathcal S_3+\mathcal S_4,
\label{eq:chandrasekhar_variational_identity}
\end{equation}
where:
\begin{align}
\mathcal I_0
&\equiv
\int_0^{R_*}
e^{(3\lambda-\nu)/2}
(P+\epsilon)r^2\xi^2\,dr,
\label{eq:inertial_integral}
\\
\mathcal S_1
&\equiv
4\int_0^{R_*}
e^{(\lambda+\nu)/2}
\frac{dP}{dr}\,
r\xi^2\,dr,
\label{eq:S1}
\\
\mathcal S_2
&\equiv
\int_0^{R_*}
e^{(\lambda+3\nu)/2}
\frac{\gamma P}{r^2}
\left(\frac{du}{dr}\right)^2dr,
\label{eq:S2}
\\
\mathcal S_3
&\equiv
\frac{8\pi G}{c^4}
\int_0^{R_*}
e^{(3\lambda+\nu)/2}
P(P+\epsilon)r^2\xi^2\,dr,
\label{eq:S3}
\\
\mathcal S_4
&\equiv
-\int_0^{R_*}
e^{(\lambda+\nu)/2}
\frac{1}{P+\epsilon}
\left(\frac{dP}{dr}\right)^2
r^2\xi^2\,dr.
\label{eq:S4}
\end{align}
These quantities coincide exactly with the volume integrals in \cite{1964chandra}.

We instead consider a pressure-confined configuration whose equilibrium
boundary satisfies:
\begin{equation}
P(R_*)=P_{\rm ext}.
\label{eq:pressure_matching_equilibrium}
\end{equation}
We assume that the pressure remains continuous across the displaced interface, such that:
\begin{equation}
\Delta P(R_*)
=
\Delta P_{\rm ext}(R_*).
\label{eq:pressure_matching_perturbed}
\end{equation}
We characterise the instantaneous response of the confining pressure by:
\begin{equation}
s
\equiv
-\frac{d\ln P_{\rm ext}(R_*)}{d\ln R_*}.
\label{eq:external_pressure_slope}
\end{equation}
For a radial displacement of the stellar boundary:
\begin{equation}
\Delta R_*=\xi_s,
\qquad
\xi_s\equiv\xi(R_*),
\end{equation}
the external-pressure perturbation is therefore:
\begin{equation}
\Delta P_{\rm ext}
=
-sP_{\rm ext}\frac{\xi_s}{R_*}.
\label{eq:external_pressure_perturbation}
\end{equation}

Because the surface pressure is now finite, the endpoint contribution in Eq.~\eqref{eq:adiabatic_operator_integration_by_parts} no longer vanishes. Using Eq.~\eqref{eq:lagrangian_pressure_perturbation}, it can be written as:
\begin{align}
\mathcal B_{\rm C}
&\equiv
-
\left[
u
e^{(\lambda+3\nu)/2}
\frac{\gamma P}{r^2}
\frac{du}{dr}
\right]_0^{R_*}
\nonumber\\
&=
\left[
r^2e^{(\lambda+\nu)/2}
\xi\Delta P
\right]_0^{R_*}.
\label{eq:boundary_term_general}
\end{align}
Regularity again removes the central contribution, leaving:
\begin{equation}
\mathcal B_{\rm C}
=
R_*^2e^{(\lambda_s+\nu_s)/2}
\xi_s\Delta P_s.
\label{eq:general_surface_term}
\end{equation}
The pressure-confined variational identity is consequently:
\begin{equation}
\omega^2\mathcal I_0
=
\mathcal S_1+\mathcal S_2+\mathcal S_3+\mathcal S_4
+\mathcal B_{\rm C}.
\label{eq:general_variational_identity}
\end{equation}
Using Eqs.~\eqref{eq:pressure_matching_equilibrium}--\eqref{eq:external_pressure_perturbation},
the boundary contribution becomes:
\begin{equation}
\mathcal B_{\rm C}
=
-sP_sR_*
e^{(\lambda_s+\nu_s)/2}
\xi_s^2,
\label{eq:surface_term_external_pressure}
\end{equation}
where $P_s=P_{\rm ext}$ is the equilibrium pressure at the stellar surface.

The integral $\mathcal I_0$ is positive for a physical equilibrium configuration. Marginal stability is therefore reached when the restoring functional on the right-hand side of Eq.~\eqref{eq:general_variational_identity} vanishes. This treatment represents the pressure confinement entirely through the response
Eq.~\eqref{eq:external_pressure_perturbation}.

\subsection{Homologous trial displacement and rearrangements}
We now perform a sequence of rearrangements that makes the Newtonian marginal-stability threshold at $\gamma=4/3$ explicit, isolates the relativistic corrections, and reveals the contribution of the modified surface boundary condition. The resulting form is also well suited to the post-Newtonian and Eddington limits considered below. We adopt the relativistic homologous trial displacement:
\begin{equation}
\xi
=
r e^{\nu/2}.
\label{eq:relativistic_homologous_displacement}
\end{equation}
We also introduce the metric notation used in the main text and by \cite{2020lionel}:
\begin{equation}
\nu=2a,
\qquad
\lambda=2b,
\end{equation}
such that:
\begin{equation}
ds^2
=
-e^{2a}(c\,dt)^2
+
e^{2b}dr^2
+
r^2d\Omega^2.
\end{equation}
For the trial displacement in Eq.~\eqref{eq:relativistic_homologous_displacement}:
\begin{equation}
u=r^3,
\qquad
\frac{du}{dr}=3r^2,
\end{equation}
and the first two contributions to the variational functional become:
\begin{align}
\mathcal S_1
&=
4\int_0^{R_*}
e^{3a+b}
\frac{dP}{dr}r^3\,dr,
\\
\mathcal S_2
&=
9\int_0^{R_*}
e^{3a+b}
\gamma Pr^2\,dr.
\end{align}
Following the rearrangement used by Haemmerl\'e, we integrate $\mathcal S_1$ by parts. Its bulk contribution then combines naturally with $\mathcal S_2$, making the factor $\gamma-4/3$ explicit:
\begin{align}
\mathcal S_1+\mathcal S_2
={}&
9\int_0^{R_*}
e^{3a+b}
\left(\gamma-\frac{4}{3}\right)
Pr^2\,dr
\nonumber\\
&
-
12\int_0^{R_*}
e^{3a+b}
\left(
\frac{da}{dr}
+
\frac{1}{3}\frac{db}{dr}
\right)
Pr^3\,dr
\nonumber\\
&
+
4e^{3a_s+b_s}P_sR_*^3.
\label{eq:first_two_terms_reduced}
\end{align}
The final term is the endpoint contribution generated by this integration by parts. It vanishes for the vacuum boundary, but must be retained here because $P_s\neq0$.

For the same trial displacement, the remaining volume terms become:
\begin{align}
\mathcal S_3
&=
\frac{8\pi G}{c^4}
\int_0^{R_*}
e^{3(a+b)}
P(P+\epsilon)r^4\,dr,
\\
\mathcal S_4
&=
-\int_0^{R_*}
e^{3a+b}
\frac{1}{P+\epsilon}
\left(\frac{dP}{dr}\right)^2
r^4\,dr.
\end{align}
Following \cite{2020lionel}, we define the four interior integrals:
\begin{align}
I_1
&\equiv
9\int_0^{R_*}
e^{3a+b}
\left(\gamma-\frac{4}{3}\right)
Pr^2\,dr,
\label{eq:I1_definition}
\\
I_2
&\equiv
-12\int_0^{R_*}
e^{3a+b}
\left(
\frac{da}{dr}
+
\frac{1}{3}\frac{db}{dr}
\right)
Pr^3\,dr,
\label{eq:I2_definition}
\\
I_3
&\equiv
\frac{8\pi G}{c^4}
\int_0^{R_*}
e^{3(a+b)}
P(P+\epsilon)r^4\,dr,
\label{eq:I3_definition}
\\
I_4
&\equiv
-\int_0^{R_*}
e^{3a+b}
\frac{1}{P+\epsilon}
\left(\frac{dP}{dr}\right)^2
r^4\,dr.
\label{eq:I4_definition}
\end{align}
In terms of the original functionals of \cite{1964chandra}:
\begin{equation}
\mathcal S_1+\mathcal S_2
=
I_1+I_2+4e^{3a_s+b_s}P_sR_*^3,
\qquad
\mathcal S_3=I_3,
\qquad
\mathcal S_4=I_4.
\label{eq:chandrasekhar_to_haemmerle_integrals}
\end{equation}
For the homologous displacement, the physical boundary contribution in Eq.~\eqref{eq:surface_term_external_pressure} becomes:
\begin{equation}
\mathcal B_{\rm C}
=
-se^{3a_s+b_s}P_sR_*^3.
\label{eq:homologous_external_boundary_term}
\end{equation}
The complete marginal-stability condition is therefore:
\begin{equation}
0
=
I_1+I_2+I_3+I_4
+
(4-s)e^{3a_s+b_s}P_sR_*^3.
\label{eq:dimensional_pressure_confined_criterion}
\end{equation}
The coefficient $4-s$ combines two distinct contributions. The term proportional to $4$ is the endpoint remainder generated when $\mathcal S_1$ is rewritten in the Haemmerl\'e form, whereas the term proportional to $-s$ is the physical response of the external pressure to the displacement of the stellar boundary.

\subsection{Post-Newtonian and Eddington expansion}
\label{sec:dimensional_PN_expansion}

We now expand the stability condition Eq.~\eqref{eq:dimensional_pressure_confined_criterion} near the Newtonian and Eddington limits. This makes the competition between gas pressure, general relativity, and external confinement explicit before introducing any dimensionless stellar variables. Near the Eddington limit, the first adiabatic exponent is:
\begin{equation}
\gamma-\frac{4}{3}
=
\frac{\beta}{6}
+
\mathcal O(\beta^2),
\qquad
\beta\equiv\frac{P_{\rm gas}}{P}.
\label{eq:eddington_gamma_expansion}
\end{equation}
The first integral therefore becomes:
\begin{equation}
I_1
=
\frac{3}{2}
\int_0^{R_*}
\beta P r^2\,dr
+
\mathcal O(\beta^2).
\label{eq:I1_dimensional_expansion}
\end{equation}

Relativistic corrections enter through the remaining three integrals. To first post-Newtonian order, the equilibrium metric derivatives are:
\begin{align}
\frac{da}{dr}
&=
\frac{GM_r}{r^2c^2}
+
\mathcal O(c^{-4}),
\label{eq:a_prime_PN}
\\
\frac{db}{dr}
&=
\frac{G}{c^2}
\left(
4\pi r\rho
-
\frac{M_r}{r^2}
\right)
+
\mathcal O(c^{-4}).
\label{eq:b_prime_PN}
\end{align}
Hence:
\begin{equation}
\frac{da}{dr}
+
\frac{1}{3}\frac{db}{dr}
=
\frac{G}{c^2}
\left(
\frac{2M_r}{3r^2}
+
\frac{4\pi}{3}r\rho
\right)
+
\mathcal O(c^{-4}).
\label{eq:metric_derivative_combination_PN}
\end{equation}
Since $I_2$ is already of post-Newtonian order, all remaining metric factors in its integrand may be evaluated in the Newtonian limit. This
gives:
\begin{equation}
I_2
=
-\frac{8G}{c^2}
\int_0^{R_*}
PM_r r\,dr
-
\frac{16\pi G}{c^2}
\int_0^{R_*}
P\rho r^4\,dr
+
\mathcal O(c^{-4}).
\label{eq:I2_dimensional_expansion}
\end{equation}
Similarly, using:
\begin{equation}
\epsilon
=
\rho c^2
+
\mathcal O(P),
\end{equation}
the third integral becomes:
\begin{equation}
I_3
=
\frac{8\pi G}{c^2}
\int_0^{R_*}
P\rho r^4\,dr
+
\mathcal O(c^{-4}).
\label{eq:I3_dimensional_expansion}
\end{equation}
For the fourth integral, Newtonian hydrostatic equilibrium:
\begin{equation}
\frac{dP}{dr}
=
-\frac{GM_r\rho}{r^2},
\label{eq:newtonian_hydrostatic_equilibrium}
\end{equation}
gives:
\begin{equation}
I_4
=
-\frac{G^2}{c^2}
\int_0^{R_*}
M_r^2\rho\,dr
+
\mathcal O(c^{-4}).
\label{eq:I4_dimensional_expansion}
\end{equation}

The sum of the four interior integrals is therefore:
\begin{align}
\sum_{i=1}^4 I_i
={}&
\frac{3}{2}
\int_0^{R_*}
\beta P r^2\,dr
-
\frac{8G}{c^2}
\int_0^{R_*}
PM_r r\,dr
\nonumber\\
&
-
\frac{8\pi G}{c^2}
\int_0^{R_*}
P\rho r^4\,dr
-
\frac{G^2}{c^2}
\int_0^{R_*}
M_r^2\rho\,dr
\nonumber\\
&
+
\mathcal O\!\left(
\beta^2,
\beta\mathcal C_*,
\mathcal C_*^2
\right),
\label{eq:sum_I_dimensional_PN}
\end{align}
where:
\begin{equation}
\mathcal C_*
\equiv
\frac{GM_*}{R_*c^2}
\end{equation}
denotes the characteristic stellar compactness.

If the mass--energy of the confining medium is neglected in the background metric, the exterior geometry at the stellar surface is
Schwarzschild. Consequently:
\begin{equation}
e^{3a_s+b_s}
=
1-\frac{2GM_*}{R_*c^2}.
\label{eq:surface_metric_dimensional}
\end{equation}
The marginal-stability condition Eq.~\eqref{eq:dimensional_pressure_confined_criterion} thus becomes:
\begin{align}
0
={}&
\underbrace{
\frac{3}{2}
\int_0^{R_*}
\beta P r^2\,dr
}_{\text{gas stabilisation}}
\nonumber\\
&
+
\underbrace{
\left[
-\frac{8G}{c^2}
\int_0^{R_*}
PM_r r\,dr
-
\frac{8\pi G}{c^2}
\int_0^{R_*}
P\rho r^4\,dr
-
\frac{G^2}{c^2}
\int_0^{R_*}
M_r^2\rho\,dr
\right]
}_{\text{relativistic destabilisation}}
\nonumber\\
&
+
\underbrace{
(4-s)P_sR_*^3
\left(
1-\frac{2GM_*}{R_*c^2}
\right)
}_{\text{pressure confinement}}
+
\mathcal O\!\left(
\beta^2,
\beta\mathcal C_*,
\mathcal C_*^2
\right).
\label{eq:dimensional_PN_stability_condition}
\end{align}
The first term is the stabilising gas-pressure correction, the following three terms are the destabilising post-Newtonian corrections, and the final term is the work associated with the confining pressure. In particular, the leading external-pressure correction is proportional to $(4-s)P_sR_*^3$, in agreement with the heuristic argument of Section~\ref{S:GR_Instability:heuristic}.

\subsection{Dimensionless hylotropic stability criterion}
\label{sec:dimensionless_hylotropic_stability}

We now rewrite the post-Newtonian stability condition in the dimensionless variables as in \cite{2020lionel}. To avoid confusing the dimensionless radial coordinate with the Lagrangian displacement $\xi(r)$ used above, we denote the former by $x$:
\begin{equation}
x=\alpha r,
\qquad
\rho=\rho_c\theta^3,
\qquad
P=P_c\psi,
\qquad
\phi=\frac{\alpha^3M_r}{4\pi\rho_c},
\label{eq:dimensionless_hylotropic_variables}
\end{equation}
where:
\begin{equation}
\alpha^2
=
\frac{\pi G\rho_c^2}{P_c},
\qquad
\sigma
\equiv
\frac{P_c}{\rho_cc^2}.
\label{eq:dimensionless_scales}
\end{equation}
Then the coordinate $x$ is identical to the dimensionless radius denoted by $\xi$ in \cite{2020lionel}. The pressure-confined surface is located at $x=x_s$, with:
\begin{equation}
\psi(x_s)=q_{\rm ext},
\qquad
q_{\rm ext}
\equiv
\frac{P_{\rm ext}}{P_c},
\qquad
\phi_s\equiv\phi(x_s).
\label{eq:dimensionless_surface_definitions}
\end{equation}
The Newtonian equations of hydrostatic equilibrium and mass continuity become:
\begin{equation}
\frac{d\psi}{dx}
=
-4\frac{\phi\theta^3}{x^2},
\qquad
\frac{d\phi}{dx}
=
x^2\theta^3.
\label{eq:dimensionless_structure_equations}
\end{equation}
Since the factors $\beta_c$ and $\sigma$ already extract the leading departures from the Eddington and Newtonian limits, respectively, all dimensionless structure integrals may be evaluated on the corresponding Newtonian equilibrium truncated at $\psi=q_{\rm ext}$.

Near the Eddington limit:
\begin{equation}
\gamma-\frac{4}{3}
=
\frac{\beta}{6}
=
\frac{\beta_c}{6}
\frac{\beta}{\beta_c}.
\end{equation}
We define the four dimensionless structure integrals:
\begin{align}
J_1
&\equiv
\int_0^{x_s}
\frac{\beta}{\beta_c}
\psi x^2\,dx,
\label{eq:J1_definition}
\\
J_2
&\equiv
\int_0^{x_s}
x\psi\phi\,dx,
\label{eq:J2_definition}
\\
J_3
&\equiv
\int_0^{x_s}
\phi^2\theta^3\,dx,
\label{eq:J3_definition}
\\
J_4
&\equiv
\int_0^{x_s}
x^4\psi\theta^3\,dx.
\label{eq:J4_definition}
\end{align}
The four dimensional integrals then reduce to:
\begin{align}
I_1
&=
\frac{3}{2}
\beta_c
\frac{P_c}{\alpha^3}
J_1,
\label{eq:I1_to_J1}
\\
I_2
&=
-16\sigma
\frac{P_c}{\alpha^3}
\left(
J_4+2J_2
\right),
\label{eq:I2_to_J}
\\
I_3
&=
8\sigma
\frac{P_c}{\alpha^3}
J_4,
\label{eq:I3_to_J4}
\\
I_4
&=
-16\sigma
\frac{P_c}{\alpha^3}
J_3.
\label{eq:I4_to_J3}
\end{align}
Their sum is therefore:
\begin{equation}
\sum_{i=1}^4 I_i
=
\frac{3P_c}{2\alpha^3}
\left[
\beta_cJ_1
-
\frac{16}{3}\sigma
\left(
J_4+4J_2+2J_3
\right)
\right].
\label{eq:sum_I_dimensionless}
\end{equation}

The surface contribution in Eq.~\eqref{eq:dimensional_pressure_confined_criterion} becomes:
\begin{equation}
(4-s)e^{3a_s+b_s}P_sR_*^3
=
(4-s)
\frac{P_c}{\alpha^3}
e^{3a_s+b_s}
q_{\rm ext}x_s^3.
\label{eq:dimensionless_surface_term}
\end{equation}
Neglecting the stress-energy of the confining medium in the background metric, the exterior geometry is Schwarzschild at the surface. Hence:
\begin{equation}
e^{3a_s+b_s}
=
1-\frac{2GM_*}{R_*c^2}
=
1-8\sigma\frac{\phi_s}{x_s}
+
\mathcal O(\sigma^2),
\label{eq:surface_metric_expansion}
\end{equation}
where:
\begin{equation}
\frac{2GM_*}{R_*c^2}
=
8\sigma\frac{\phi_s}{x_s}.
\label{eq:dimensionless_compactness}
\end{equation}
Multiplying the marginal-stability condition by
$2\alpha^3/(3P_c)$ gives:
\begin{align}
0
={}&
\beta_cJ_1
-
\frac{16}{3}\sigma
\left(
J_4+4J_2+2J_3
\right)
\nonumber\\
&
+
\frac{2}{3}(4-s)
q_{\rm ext}x_s^3
\left(
1-8\sigma\frac{\phi_s}{x_s}
\right).
\label{eq:pressure_confined_criterion_J4}
\end{align}

The integral $J_4$ is not independent. Using mass continuity:
\begin{align}
J_4
&=
\int_0^{x_s}
x^2\psi
\frac{d\phi}{dx}\,dx
\nonumber\\
&=
\left[
x^2\psi\phi
\right]_0^{x_s}
-
\int_0^{x_s}
\phi
\frac{d}{dx}
\left(
x^2\psi
\right)dx.
\label{eq:J4_integration_by_parts}
\end{align}
The central endpoint vanishes by regularity, while the surface endpoint is:
\begin{equation}
\left[
x^2\psi\phi
\right]_{x=x_s}
=
q_{\rm ext}x_s^2\phi_s.
\end{equation}
Using Eq.~\eqref{eq:dimensionless_structure_equations}, we obtain:
\begin{equation}
J_4
=
q_{\rm ext}x_s^2\phi_s
-
2J_2
+
4J_3.
\label{eq:J4_pressure_boundary}
\end{equation}

Substituting Eq.~\eqref{eq:J4_pressure_boundary} into
Eq.~\eqref{eq:pressure_confined_criterion_J4} gives:
\begin{align}
0
={}&
\beta_cJ_1
-
\frac{32}{3}\sigma
\left(
J_2+3J_3
\right)
+
\frac{2}{3}(4-s)
q_{\rm ext}x_s^3
\nonumber\\
&
-
\frac{16}{3}(5-s)
\sigma q_{\rm ext}x_s^2\phi_s.
\label{eq:pressure_confined_criterion_reduced}
\end{align}
The final term combines the finite-pressure endpoint generated when $J_4$ is eliminated with the post-Newtonian correction to the metric factor multiplying the surface work. Relative to the leading external-pressure contribution, its magnitude is:
\begin{equation}
\left|
\frac{
-\frac{16}{3}(5-s)
\sigma q_{\rm ext}x_s^2\phi_s
}{
\frac{2}{3}(4-s)
q_{\rm ext}x_s^3
}
\right|
=
\left|
\frac{5-s}{4-s}
\right|
\frac{2GM_*}{R_*c^2}.
\label{eq:mixed_surface_term_ratio}
\end{equation}
It is therefore suppressed by the stellar compactness and may be neglected at leading order for a weakly relativistic star, provided $s$ is not close to $4$. The stability criterion used for the numerical calculations is therefore:
\begin{equation}
0
=
\beta_cJ_1
-
\frac{32}{3}\sigma
\left(
J_2+3J_3
\right)
+
\frac{2}{3}(4-s)
q_{\rm ext}x_s^3,
\label{eq:pressure_confined_criterion_leading}
\end{equation}
up to corrections of order $\sigma q_{\rm ext}$. 

\bibliographystyle{aasjournal}
\bibliography{main}

\end{document}